\documentclass[aps,pre,superscriptaddress,amsmath,amssymb,superscriptaddress,twocolumn,floatfix,showpacs,longbibliography]{revtex4-1}
\usepackage{graphicx}
\usepackage{float}
\usepackage[tight,footnotesize]{subfigure}
\usepackage{color}

\begin{document}
\title{Jamming transitions induced by an attraction in pedestrian flow} 

\author{Jaeyoung Kwak}
\email{jaeyoung.kwak@aalto.fi}
\affiliation{Department of Built Environment, Aalto University, Espoo, Finland}
\author{Hang-Hyun Jo}
\affiliation{Asia Pacific Center for Theoretical Physics, Pohang, Republic of Korea}
\affiliation{Department of Physics, Pohang University of Science and Technology, Pohang, Republic of Korea}
\affiliation{Department of Computer Science, Aalto University, Espoo, Finland}
\author{Tapio Luttinen}
\affiliation{Department of Built Environment, Aalto University, Espoo, Finland}
\author{Iisakki Kosonen}
\affiliation{Department of Built Environment, Aalto University, Espoo, Finland}
\date{\today}

\begin{abstract}
We numerically study jamming transitions in pedestrian flow interacting with an attraction, mostly based on the social force model for pedestrians who can join the attraction. We formulate the joining probability as a function of social influence from others, reflecting that individual choice behavior is likely influenced by others. By controlling pedestrian influx and the social influence parameter, we identify various pedestrian flow patterns. For the bidirectional flow scenario, we observe a transition from the free flow phase to the freezing phase, in which oppositely walking pedestrians reach a complete stop and block each other. On the other hand, a different transition behavior appears in the unidirectional flow scenario, i.e., from the free flow phase to the localized jam phase and then to the extended jam phase. It is also observed that the extended jam phase can end up in freezing phenomena with a certain probability when pedestrian flux is high with strong social influence. This study highlights that attractive interactions between pedestrians and an attraction can trigger jamming transitions by increasing the number of conflicts among pedestrians near the attraction. In order to avoid excessive pedestrian jams, we suggest suppressing the number of conflicts under a certain level by moderating pedestrian influx especially when the social influence is strong.
\end{abstract}


\maketitle

\section{Introduction}
Collective dynamics of many-body systems has attracted much attention in the fields of statistical physics and its neighboring disciplines. As for the examples, one finds the collective motions of particles~\cite{Vicsek_PRL1995}, vehicles~\cite{Helbing_RMP2001}, pedestrians~\cite{Helbing_PRE1995}, and animals~\cite{Couzin_ProcB2002}. This subject has been studied by modeling a set of individual behavioral rules in order to quantify emergent collective patterns from interactions among individuals. Based on this approach, various interesting collective behaviors have been identified such as the coherent state in highway traffic~\cite{Helbing_Nature1998} and lane formation in pedestrian flow~\cite{Helbing_PRE1995}. These collective behaviors are interesting not only because they arise without any external controls but also because they improve the efficiency of traffic flow. However, for the density of particles above a certain level, the interactions among individuals may cause jamming transitions that reduce the traffic flow efficiency~\cite{Sugiyama_NJP2008, Nakayama_NJP2009, Tadaki_NJP2013}. Jamming transitions have generated considerable research interest, not only because of their relevance to collective dynamics including the clogging effect in granular flow~\cite{Zuriguel_PRL2011} and the faster-is-slower effect in pedestrian evacuations~\cite{Helbing_Nature2000}, but also for practical applications such as monitoring congestion on freeways~\cite{Kerner_TrC2004, Kerner_2004} and developing adaptive cruise control strategies~\cite{Kesting_TrC2008}.

In order to understand jamming transitions and related phenomena in pedestrian flow, experimental studies have been performed for unidirectional~\cite{Seyfried_JSTAT2005, Zhang_JSTAT2011} and bidirectional flow scenarios~\cite{Kretz_JSTAT2006_10a, Feliciani_PRE2016, Zhang_JSTAT2012}. Seyfried~\textit{et al.}~\cite{Seyfried_JSTAT2005} and Zhang~\textit{et al.}~\cite{Zhang_JSTAT2011, Zhang_JSTAT2012} studied the shape of fundamental diagrams based on various pedestrian flow experiments. For different sizes of two oppositely walking pedestrian groups, Kretz~\textit{et al.}~\cite{Kretz_JSTAT2006_10a} examined the characteristics of bidirectional flow by looking into passing times, walking speeds, fluxes, and lane formation. Feliciani and Nishinari~\cite{Feliciani_PRE2016} investigated the lane formation process based on experiment data of different directional split in bidirectional flow. Pedestrian flow through bottlenecks has been also actively studied~\cite{Kretz_JSTAT2006_10b, Hoogendoorn_TrSci2005, Seyfried_TrSci2009}. Those bottleneck studies analyzed the influence of bottleneck width on pedestrian flow including bottleneck capacity, time headways, and total times to flee all the pedestrians from the bottleneck. Up to now, most of pedestrian bottleneck studies have been performed for static bottlenecks, meaning that the bottlenecks are at fixed locations and their size does not change over time. 

Jamming transitions in pedestrian flow have also been investigated for various situations based on numerical simulations. With the lattice gas model, Muramatsu~\textit{et al.}~\cite{Muramatsu_PhysicaA1999} studied jamming transitions as a function of pedestrian density in bidirectional flow and observed a freezing transition for high pedestrian density in a straight corridor. Later, Tajima~\textit{et al.}~\cite{Tajima_PhysicaA2001} identified a jamming transition from free flow to saturated flow at a critical density. Above the critical density, the pedestrian flow rate stays constant against increasing density, defining the saturated flow rate. They also presented the scaling behavior of the saturated flow rate and critical density depending on the width of the bottleneck and corridor. For an evacuation scenario, Helbing~\textit{et al.}~\cite{Helbing_Nature2000} found that an arch-like blocking appears in front of an exit which significantly increases the evacuation time. In another study, they reported that a noise term in the equation of pedestrian motion can reproduce a freezing phenomenon in which pedestrian flow reaches a complete stop~\cite{Helbing_PRL2000}. Recently, Yanagisawa investigated the influence of memory effect on bidirectional flow in a narrow corridor. When the memory-loss rate is above a certain value, oppositely walking pedestrians fail to avoid encountering each other, leading to clogging~\cite{Yanagisawa_CDyn2016}.

A considerable amount of literature has reported jamming transitions in the flow of pedestrians walking from one point to another. In addition, previous studies provided narrative descriptions of the case interacting with attractions such as shop displays and public events. For instance, Goffman~\cite{Goffman_1971} described that window shoppers act like obstructions to passersby on streets when they stop to check store displays. Those shoppers can further interfere with other pedestrians when the shoppers enter and leave the stores. In another study, Gipps and Marksj\"{o}~\cite{Gipps_MCS1985} stated that an attraction in a pedestrian facility can attract nearby pedestrians, and such an attraction may impede pedestrian traffic especially during peak periods. 

Although it has been well recognized that an attraction can trigger pedestrian jams, little attention has been paid to characterize the dynamics of their jamming transitions. In pedestrian facilities, pedestrians can see the attractions and might shift their attention towards the attractions. If the attractions are tempting enough, a fair number of pedestrians gather around the attractions, forming attendee clusters. In our previous studies~\cite{Kwak_PRE2013, Kwak_PLOS2015}, we characterized collective patterns of attendee clusters, and investigated how such various patterns can emerge from attractive interactions between pedestrians and attractions. Nevertheless, little is known about how an attendee cluster can contribute to jamming transitions in pedestrian flow. It is apparent that if a large attendee cluster exists near an attraction, passersby are forced to walk through the reduced available space. Consequently, the attendee cluster is acting as a pedestrian bottleneck for passersby. The flow through the bottleneck can show transitions from free flow to jamming states and may end in gridlock. The jamming transitions near an attraction will be the subject of this paper. In the following, we investigate jamming patterns induced by the attraction and understand the transitions at a microscopic level.

By means of numerical simulations, we characterize jamming transitions in pedestrian flow interacting with an attraction. The simulation model and its setup are explained in Sec.~\ref{sec:Model}. Then we analyze the spatio-temporal patterns of jamming transitions induced by an attraction and summarize the results with phase diagrams, as shown in Sec.~\ref{sec:Results}. We provide microscopic understanding of the jamming transitions by mainly looking into conflicts among pedestrians. Finally, we discuss the findings of this study in Sec.~\ref{sec:Conclusion}.

\section{Model}
\label{sec:Model}
Following the work of Helbing and Moln\'{a}r~\cite{Helbing_PRE1995}, we describe the motion of pedestrian $i$ with the following equation:
\begin{eqnarray}\label{eq:EoM}
	\frac{\mathrm{d} \vec{v}_i(t)}{\mathrm{d} t} = \frac{v_d\vec{e}_{i}-\vec{v}_{i}}{\tau} + \sum_{j\neq i}^{ }{\vec{f}_{ij}} + \sum_{B}^{ }{\vec{f}_{iB}}.
\end{eqnarray}
The first term on the right hand-side is the driving force term indicating that pedestrian $i$ adjusts walking velocity $\vec{v}_{i}$ in order to achieve a desired walking speed $v_d$ along with the desired walking direction vector $\vec{e}_{i}$. Here $\vec{e}_{i}$ is a unit vector pointing to the direction in which pedestrian $i$ wants to move. The relaxation time $\tau$ controls how quickly the pedestrian adapts one's velocity to the desired velocity. The repulsive force terms $\vec{f}_{ij}$ and $\vec{f}_{iB}$ reflect the pedestrian's collision avoidance behavior against another pedestrian $j$ and the boundary $B$, respectively. In this section, the details of Eq.~(\ref{eq:EoM}) and the numerical simulation setup are explained. 

\subsection{Desired walking speed}
\label{sec:v_d}
Previous studies have reported that preventing excessive overlaps among pedestrians is important to provide better representation of pedestrian stopping behavior, which often triggers jams. Parisi~\textit{et al.}~\cite{Parisi_PhysicaA2009} introduced the respect area, which reserves a space on the order of pedestrian radius, in order to suppress overlapping among pedestrians. Later, Chraibi~\textit{et al.}~\cite{Chraibi_PRE2015} proposed an interpersonal repulsion model that can prevent overlapping in one dimensional pedestrian flow. In their models, the driving force term becomes inactive when a pedestrian does not have enough room for stride. Inspired by those studies, we postulate that the desired speed $v_d$ is an attainable speed of pedestrian $i$ depending on the available walking space in front of the pedestrian, 
\begin{eqnarray}\label{eq:desired_speed}
    v_d = \min\{v_0, d_{ij}/T_c\},
\end{eqnarray}
where $v_0$ is a comfortable walking speed and $d_{ij}$ is the distance between pedestrian $i$ and the first pedestrian $j$ encountering with pedestrian $i$ in the course of $\vec{v}_i$. Time-to-collision $T_c$ represents how much time remains for a collision of two pedestrians $i$ and $j$. Further details of $T_c$ are given in Appendix~\ref{sec:T_c}.

\subsection{Collision avoidance behavior}
\label{sec:collision_avoidance}
The collision avoidance behavior is modeled with $\vec{f}_{ij}$ and $\vec{f}_{iB}$. Previous studies~\cite{Helbing_PRE1995, Johansson_ACS2007, Kwak_PRE2013, Kwak_PLOS2015} specified the interpersonal repulsion term $\vec{f}_{ij}$ as a derivative of repulsive potential with respect to $\vec{d}_{ij} \equiv \vec{x_i} - \vec{x_j}$. It is given as 
\begin{eqnarray}\label{eq:repulsive1}
	\vec{f}_{ij} = -\nabla_{\vec{d}_{ij}}\left[ C_pl_p\exp\left(-\frac{b_{ij}}{l_p}\right) \right] \omega_{ij}.
\end{eqnarray}
Here, $C_p$ and $l_p$ are the strength and range of the interpersonal repulsion, and $b_{ij} = \frac{1}{2}\sqrt{(\|\vec{d}_{ij}\|+\|\vec{d}_{ij}-\vec{y}_{ij}\|)^{2}-\|\vec{y}_{ij}\|^{2}}$ is the effective distance between pedestrians $i$ and $j$ by assuming their relative displacement $\vec{y}_{ij}\equiv (\vec{v}_{j}-\vec{v}_{i})\Delta t_s$ with the stride time $\Delta t_s$~\cite{Johansson_ACS2007}. The anisotropic function $\omega_{ij}$ represents the directional sensitivity to pedestrian $j$, 
\begin{eqnarray}\label{eq:anisotropic}
	\omega_{ij} = \lambda_{ij} + (1-\lambda_{ij})\frac{1+\cos~\phi_{ij}}{2}, %
\end{eqnarray}
where $0 \leq \lambda_{ij} \leq 1$ is pedestrian $i$'s minimum anisotropic strength against pedestrian $j$. In addition, the angle $\phi_{ij}$ is measured between the velocity vector of the pedestrian $i$, $\vec{v}_{i}$, and relative location of pedestrian $j$ with respect to pedestrian $i$, $\vec{d}_{ji} = \vec{x}_j - \vec{x}_i$.

The boundary repulsion is given as $\vec{f}_{iB} = C_b\exp[(r_{i}-{d_{iB}})/{l_{b}}]\vec{e}_{iB}$, where $d_{iB}$ is the perpendicular distance between pedestrian $i$ and wall $B$, and $\vec{e}_{iB}$ is the unit vector pointing from the wall $B$ to the pedestrian $i$. The strength and the range of repulsive interaction from boundaries are denoted by $C_b$ and $l_b$, respectively. 

\subsection{Joining behavior}
\label{sec:joining_behavior}
It has been widely believed that individual choice behavior can be influenced by the choice of other individuals. For instance, previous studies on stimulus crowd effects reported that a pedestrian is more likely to shift his attention towards the crowd as its size grows~\cite{Milgram_JPSP1969, Gallup_PNAS2012}. This belief is also generally accepted in the marketing area, which can be interpreted that having more visitors in a store can attract more pedestrians to the store~\cite{Bearden_JCR1989, Childers_JCR1992}. It is also suggested that the sensitivity to others' choice is different for different places, time-of-day, and visitors' motivation~\cite{Gallup_PNAS2012, Kaltcheva_JMkt2006}. Based on those studies~\cite{Milgram_JPSP1969, Gallup_PNAS2012, Bearden_JCR1989, Childers_JCR1992, Kaltcheva_JMkt2006}, we assume that an individual decides whether to visit an attraction based on the number of pedestrians attending the attraction. The sensitivity to others' choice can be represented as the social influence parameter $s$. As suggested by Ref.~\cite{Kwak_PLOS2015}, we formulate the probability of joining an attraction $P_a$ by the analogy with sigmoidal choice rule~\cite{Milgram_JPSP1969, Gallup_PNAS2012, Nicolis_PRL2013}, 
\begin{eqnarray}\label{eq:P_a}
	P_{a} = \frac{s(N_{a}+K_a)}{({N_{0}+K_0})+s({N_{a}+K_a})}.
\end{eqnarray}
Here, $N_{a}$ and $N_{0}$ are the number of pedestrians who have already joined and that of the pedestrians not stopping by the attraction, respectively. In order to prevent the indeterminate case of Eq.~(\ref{eq:P_a}), we set $K_a$ and $K_0$ as baseline values for $N_a$ and $N_0$. The social influence parameter $s > 0$ can be also understood as pedestrians' awareness of the attraction. According to previous studies~\cite{Milgram_JPSP1969, Gallup_PNAS2012, Kaltcheva_JMkt2006}, we assume that the strength of social influence can be different for different situations and can be controlled in the presented model. Once an individual has joined an attraction, the individual will then stay near the attraction for an exponentially distributed time with an average of $t_{d}$~\cite{Helbing_PRE1995, Kwak_PLOS2015, Gallup_PNAS2012}. After the duration of visit, one leaves the attraction and continues walking towards one's initial destination, not visiting the attraction again. 

\subsection{Steering behavior of passersby}
\label{sec:steering_behavior}
While attracted pedestrians are joining the attraction according to Eq.~(\ref{eq:P_a}), passersby are the pedestrians who are not interested in the attraction, thus they do not visit the attraction. In this study, we assume passersby aim at smoothly bypassing an attendee cluster near the attraction while walking towards their destination. 

Various approaches are available for modeling pedestrian steering behavior, including the pedestrian stream model~\cite{Hughes_TrB2002}, the Voronoi diagram based approach~\cite{Xiao_TrC2016}, and dynamic floor field models~\cite{Kretz_JSTAT2009, Hartmann_NJP2010, Kneidl_TrC2013, Hartmann_CiCP2014}. Among these approaches, the dynamic floor field models have been widely applied to model pedestrian steering behavior. Similar to cellular automata based pedestrian models~\cite{Burstedde_PhysicaA2001, Kirchner_PhysicaA2002}, the dynamic floor field models discretize the pedestrian walking space into grids on the order of pedestrian size. In line with the eikonal equation, the walking speed at each grid point is assumed to be inversely proportional to the derivative of the expected travel time function. That is, a pedestrian standing at the grid point is going to walk in a direction minimizing the expected travel time to a destination. The expected travel time is updated based on the local pedestrian density at every time step. Analogously to wave propagation in fluids, the expected travel time is calculated along a pathway from the destination to the grid point, inferring that pedestrians can plan ahead to take a pathway offering the shortest travel time. 
That is, the dynamic floor field models consider that pedestrians walk along the fastest way to the destination. Previous studies demonstrated that using the models can significantly improve pedestrian steering behavior in numerical simulations~\cite{Kretz_JSTAT2009, Hartmann_NJP2010, Kneidl_TrC2013}. However, the approach is computationally expensive mainly due to the calculation of the local density for almost every time step. 

For computational efficiency, we employ streamline approach to steer passersby between boundaries of the corridor. Similar to the potential flow in fluid dynamics~\cite{Batchelor_2000, Anderson_2010} and the pedestrian stream model~\cite{Hughes_TrB2002}, streamlines are used to represent plausible trajectories of particles smoothly bypassing obstacles. For a location $z = (x, y)$, the pedestrian velocity components in the $x$- and $y$-directions are expressed as partial derivatives of a streamline function. In this study, the streamline function is formulated as a function of attendee cluster size, so that passersby can detour around an attendee cluster near the attraction. The attendee cluster size is measured at every time step. See Appendix~\ref{sec:streamline} for more details. According to the streamline approach, pedestrians decide their walking direction in response to immediate changes around them rather than based on a prediction of travel time to the destination. In contrast to the dynamic floor field models, the streamline approach is computationally efficient because it does not require one to compute the local density at every time step.

\subsection{Numerical simulation setup}
\label{sec:stetup}
Each pedestrian is modeled by a circle with radius $r_i = 0.2$~m. Pedestrians move in a corridor of length $L = 60$~m and width $W = 4$~m in the horizontal direction. An attraction is placed at the center of the lower wall. Pedestrians move with comfortable walking speed $v_0 = 1.2$~m/s and with relaxation time $\tau = 0.5$~s, and their speed cannot exceed $v_{\rm max} = 2.0$~m/s. The parameters of the repulsive force terms are given based on previous works: $C_p = 3$, $l_p = 0.3$, $\Delta t_s = 2.5$, $C_b = 6$, and $l_b = 0.3$~\cite{Helbing_PRE1995, Johansson_ACS2007, Kwak_PRE2013, Kwak_PLOS2015, Zanlungo_PLOS2012}. The minimum anisotropic strength $\lambda_{ij}$ is set to 0.25 for attendees near the attraction and 0.5 for others, yielding that the attendees exert smaller repulsive force on others than passersby do. Consequently, the attendees can stay closer to the attraction while being less disturbed by the passersby. 

The social force model in Eq.~(\ref{eq:EoM}) is updated for each simulation time step $\Delta t = 0.05$~s. Following previous studies~\cite{Xu_IEEE2010, Zanlungo_EPL2011, Zanlungo_PRE2014}, we employ the first-order Euler method for numerical integration of Eq.~(\ref{eq:EoM}). We refer the readers to Appendix~\ref{sec:numerical_integration} for further details. We note that $\Delta t = 0.05$~s yields good results in our numerical simulations without excessive overlaps among pedestrians. This is possible because pedestrians reduce their desired waking speed when they encounter other pedestrians in a course of collision. Smaller values of $\Delta t$ can be selected for better resolution of pedestrian trajectories~\cite{Koster_PRE2013}. Based on Eq.~(\ref{eq:P_a}), the joining probability is updated for every 0.05~s. For convenience, the evaluation time of Eq.~(\ref{eq:P_a}) is the same value as $\Delta t$. As pointed out by Ref~\cite{Heliovaara_PRE2013}, there is no established value for update frequency of pedestrian decision. An individual evaluates the joining probability when one can perceive the attraction $10$~m ahead. If the individual decides to join the attraction, then the desired direction vector $\vec e_i$ is changed from $\vec e_{i, 0}$ into $\vec e_{Ai}$. Here, $\vec e_{i, 0}$ and $\vec e_{Ai}$ are unit vectors indicating the initial desired walking direction of pedestrian $i$ and pointing from pedestrian $i$ to the attraction, respectively. For simplicity, $K_a$ and $K_0$ are set to be $1$, meaning that both options are equally attractive when the individual would see nobody within $10$~m from the center of the attraction. An individual $i$ is counted as an attending pedestrian if pedestrian $i$'s efficiency of motion $E_{i} \equiv (\vec{v}_{i} \cdot \vec{e}_{i, 0})/{v_0}$ is lower than $0.05$ within a range of $1$~m from the boundary of the attendee cluster. The individual efficiency of motion $E_i$ indicates how much the driving force contributes to pedestrian $i$'s progress towards the destination with a range from $0$ to $1$~\cite{Helbing_PRL2000, Kwak_PRE2013}. We assumed that $E_i = 0.05$ is tolerant enough to distinguish a weak motion of attendees near an attraction from actively walking pedestrians not visiting the attraction. The average duration of visiting an attraction $t_d$ is set to be $30$~s.

For our numerical simulations, a straight corridor will be considered to study pedestrian jams induced by an attraction. In the straight corridor, one can consider two possible patterns of flow, i.e., bidirectional and unidirectional flows. In the bidirectional flow, one half of the population is walking towards the right boundary of the corridor from the left, and the opposite direction for the other half. In the unidirectional flow, all pedestrians are entering the corridor through the left boundary and walking towards the right. 

An open boundary condition is employed in order to continuously supply pedestrians to the corridor. By doing so, pedestrians can enter the corridor regardless of the number of leaving pedestrians. Pedestrians are inserted at random places on either side of the corridor without overlapping with other nearby pedestrians and boundaries. The number of pedestrians in the corridor is associated with the pedestrian influx $Q$, i.e., the arrival rate of pedestrians entering the corridor. The unit of $Q$ is indicated by P/s, which stands for pedestrians per second. Based on previous studies~\cite{May_1990, Luttinen_TRR1999}, the pedestrian interarrival time is assumed to follow a shifted exponential distribution. That is, pedestrians are entering the corridor independently and their arrival pattern is not influenced by that of others. The minimum headway is set to $0.4$~s between successive pedestrians entering the corridor, which is large enough to prevent overlaps between arriving pedestrians. 

\section{Results and Discussion}
\label{sec:Results}
\subsection{Jam patterns in bidirectional flow}
\label{subsec:bidirectional}

\begin{figure}[!t]
	\centering
	\includegraphics[width=\columnwidth]{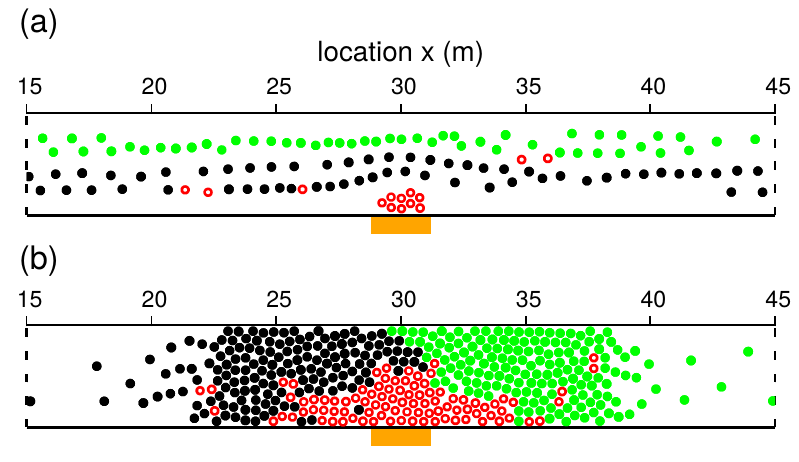}\vspace{-0.3cm}
	\caption{(Color online) Representative snapshots of different passerby flow patterns, showing a section of $30$~m in the center of the corridor, i.e., $15\text{~m} \leq x \leq 45\text{~m}$. The attraction, depicted by an orange rectangle, is located at the center of the lower wall with open boundary conditions in the horizontal direction. Closed black and green circles indicate passersby walking to the right and to the left, respectively. Open red circles depict pedestrians attracted by the attraction. In bidirectional flow, we can observe (a) the free flow phase in the case of $Q = 4$~P/s and $s = 0.5$, in which passersby can walk towards their destinations without being interrupted by the cluster of attracted pedestrians, and (b) the freezing phase in the case of $Q = 4$~P/s and $s = 1$, where oppositely walking pedestrians stopped because they block each other. Note that P/s stands for pedestrians per second, being the unit of pedestrian flux $Q$.}
	\label{fig:snapshot_bi} 
\end{figure}

Our simulation results show different patterns of pedestrian motion depending on influx $Q$ and social influence parameter $s$. The free flow phase appears when both $Q$ and $s$ are small. From Fig.~\ref{fig:snapshot_bi}(a), one can observe that passersby walk towards their destinations without being interrupted by the cluster of attracted pedestrians. Passersby walking to the right form lanes in lower part of the corridor while the upper part of the corridor is occupied by passersby walking to the left. This spatial segregation appears as a result of the lane formation process which has been reported in previous studies~\cite{Helbing_PRE1995, Kretz_JSTAT2006_10a, Zhang_JSTAT2012, Feliciani_PRE2016}. Simultaneously, the attracted pedestrians form a stable cluster near the attraction. If both $Q$ and $s$ are large, we can see freezing phase, in which oppositely walking pedestrians reach a complete stop because they block each other, as shown in Fig.~\ref{fig:snapshot_bi}(b). 

To quantify spatio-temporal patterns of pedestrian flow, we measure the local efficiency $E(x, t)$ for a given time $t$ and segment $x$ in the horizontal direction: 
\begin{eqnarray}\label{eq:efficiency_avg}
	E(x, t) = \frac{1}{|N(x, t)|} \sum_{i\in N(x, t)}{E_i}.
\end{eqnarray}
Here $N(x, t)$ is the set of passersby in a $1$~m long segment $x$ at time $t$. The individual efficiency of motion $E_i = (\vec{v}_{i} \cdot \vec{e}_{i, 0})/{v_0}$ can be understood as a normalized speed of pedestrian $i$ in the horizontal direction. The local efficiency $E(x, t)$ indicates how fast passersby in segment $x$ progress towards their destination at time $t$. If $|N(x, t)| = 0$, we set to $E(x,t) = 1$ inferring that the passersby can walk with their comfortable speed $v_0$ if they are in the segment $x$ at time $t$. Thus, $E(x,t) = 1$ indicates that the passersby can freely walk without reducing their speed, while $E(x,t) = 0$ implies that the passersby have reached a standstill. 

\begin{figure}[!t]
	\centering
	\includegraphics[width=\columnwidth]{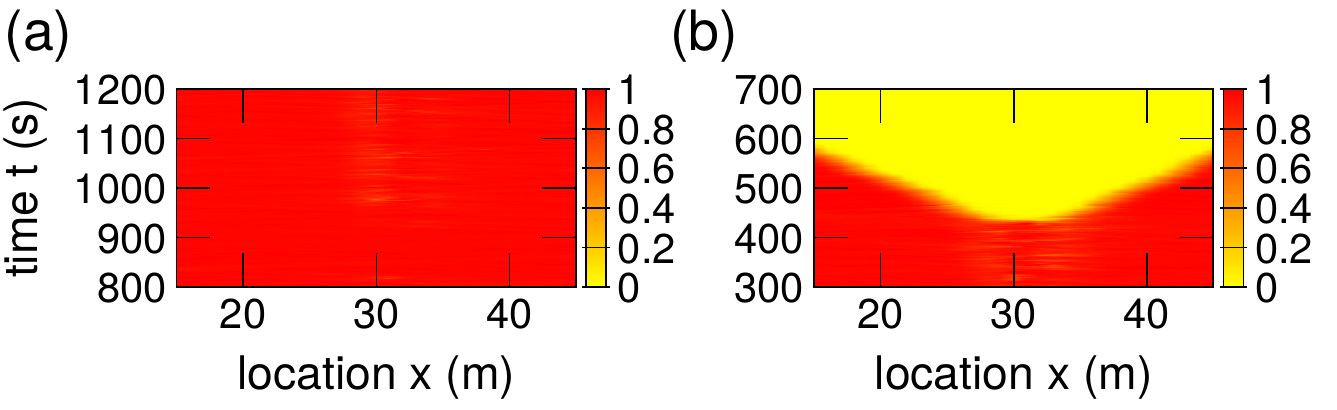}\vspace{-0.3cm}
	\caption{(Color online) Local efficiency for a given time $t$ and location $x$, $E(x, t)$, for different pedestrian flow patterns in bidirectional flow. The attraction is located at $x = 30$~m, at the center of the corridor. Light yellow and dark red colors indicate lower and higher values, respectively. (a) Free flow phase with $Q = 4$~P/s and $s = 0.5$, and (b) Freezing phase with $Q = 4$~P/s and $s = 1$. Note that the spatio-temporal representation of local efficiency is dark red for (a) $t < 800$~s and $t > 1200$~s and (b) $t < 300$~s.}
	\label{fig:spatiotemporal_bi}
\end{figure}

Figure~\ref{fig:spatiotemporal_bi} shows the corresponding spatio-temporal representation of different pedestrian flow patterns. As shown in Fig.~\ref{fig:spatiotemporal_bi}(a), in the free flow phase, the local efficiency $E(x, t)$ is almost $1$ over the stationary state period, meaning that all the passersby walk with their comfort speed. In the freezing phase, the local efficiency $E(x, t)$ suddenly becomes zero near the attraction around at $t = 450$~s, and then the low efficiency area expands to the left and right boundaries of the corridor in the course of time [see Fig.~\ref{fig:spatiotemporal_bi}(b)]. 

\begin{figure}[!t]
	\centering
	\begin{tabular}{cc}
		\includegraphics[width=.49\columnwidth]{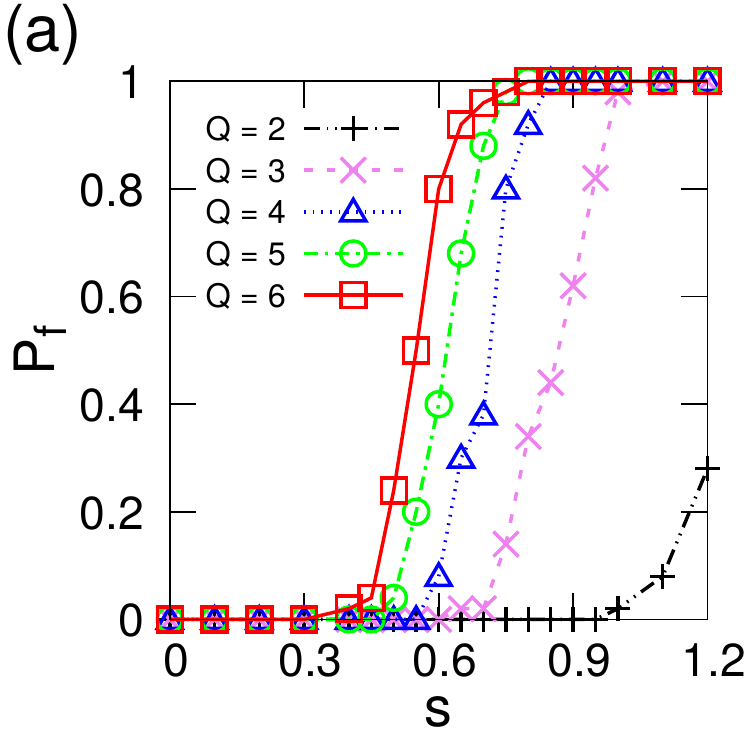}&
		\includegraphics[width=.49\columnwidth]{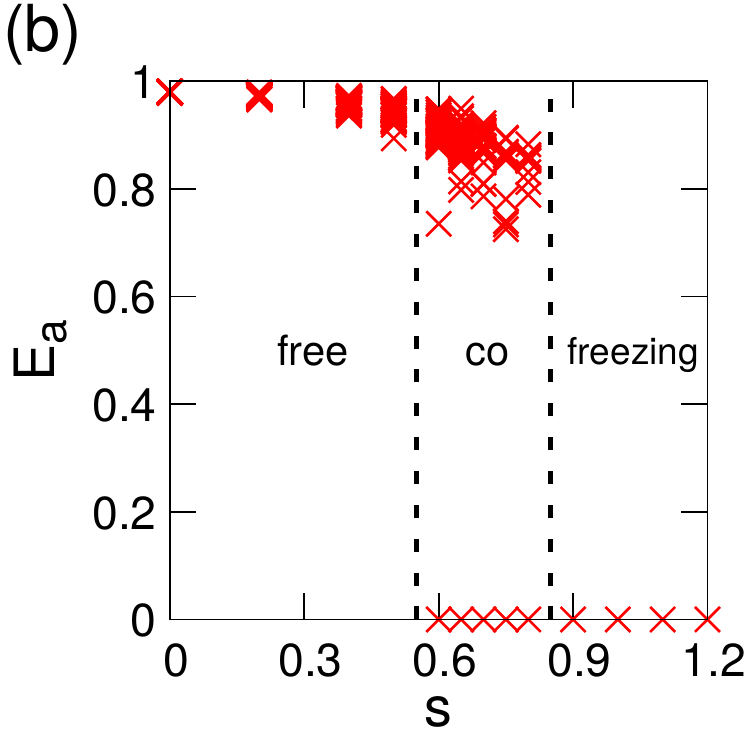}\\
		\includegraphics[width=.49\columnwidth]{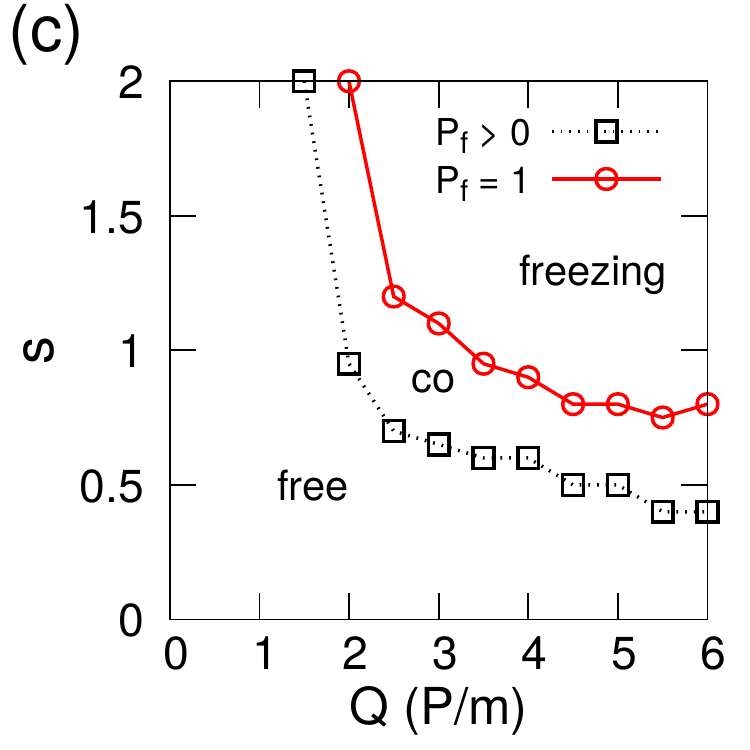}\vspace{-0.3cm}
	\end{tabular}
	\caption{(Color online) (a) Freezing probability $P_f$ as a function of influx $Q$ and social influence parameter $s$ in bidirectional flow. Different symbols represent different values of $Q$. (b) Stationary state average of local efficiency near the attraction $E_{a}$ against $s$ in bidirectional flow with $Q = 4$~P/s. Each point depicts a value of $E_{a}$ obtained from one simulation run. Here, free, freezing, and co indicate free flow phase, freezing phase, and coexisting phase, respectively. In coexisting phase, one can observe freezing phenomena with a certain probability $P_f$. (c) Phase diagram summarizing the numerical results of bidirectional flow. The parameter space of pedestrian influx $Q$ and social influence parameter $s$ is divided into different phases by means of $P_f$ for bidirectional flow.}
	\label{fig:PhaseDiagram_bi} 
\end{figure}

Next, we identify the freezing phase by means of cumulative throughput at $x = 30$~m, according to Ref.~\cite{Daganzo_1997}. If the cumulative throughput does not change for $120$~s, it infers the appearance of the freezing phenomenon. We obtain the freezing probability $P_f$ by counting the occurrence of freezing phenomena over 50 independent simulation runs for each parameter combination $(Q,\ s)$. The freezing probability $P_f$ tends to increase as $Q$ and $s$ increase, see Fig.~\ref{fig:PhaseDiagram_bi}(a). For small value of $Q \leq 1.4$~P/s, $P_f$ is zero up to $s = 2$, indicating that the freezing phenomenon is not observable. We classify parameter combinations of $(Q, s)$ yielding $P_f = 0$ as the free flow phase and $P_f = 1$ for the freezing phase. We call the parameter space between the envelopes of $P_f = 0$ and $P_f = 1$ as the coexisting phase, noting that both phases can appear depending on random seeds in the numerical simulations. 

In order to further quantify different phases, we calculate stationary state average of local efficiency, $E(x)$, in the vicinity of the attraction. Here, $E(x)$ is given as
\begin{eqnarray}\label{eq:E_x}
	E(x) = \langle E(x, t) \rangle,
\end{eqnarray}
where $\langle\cdot\rangle$ represents the average obtained from a simulation run after reaching the stationary state. We select a section of $27\text{~m} \leq x \leq 33\text{~m}$ to evaluate the stationary state average value in the vicinity of the attraction, and the minimum value of $E(x)$ is denoted by $E_{a}$. Note that $E_{a}$ is selected in a way to reflect the largest possible efficiency drop in the section. Figure~\ref{fig:PhaseDiagram_bi}(b) presents $E_{a}$ against $s$ in the case of $Q = 4$~P/s. One can observe that $E_{a}$ is almost $1$ for $s < 0.6$, depicting the free flow phase. For $0.6 \leq s \leq 0.8$, some data points of $E_{a}$ are positive while others are zero. That is, two distinct pedestrian flow patterns can be observed for the same value of $s$ depending on random seeds. For $s > 0.8$, $E_{a}$ is always zero, corresponding to the freezing phase. 

Figure~\ref{fig:PhaseDiagram_bi}(c) summarizes numerical results of phase characterizations. The parameter space of pedestrian influx $Q$ and social influence parameter $s$ is divided into different phases by means of $P_f$. In the coexisting phase, one can observe freezing phenomena with a certain probability $P_f$.

\subsection{Jam patterns in unidirectional flow}
\label{subsec:unidirectional}

\begin{figure}[!t]
	\centering
	\includegraphics[width=\columnwidth]{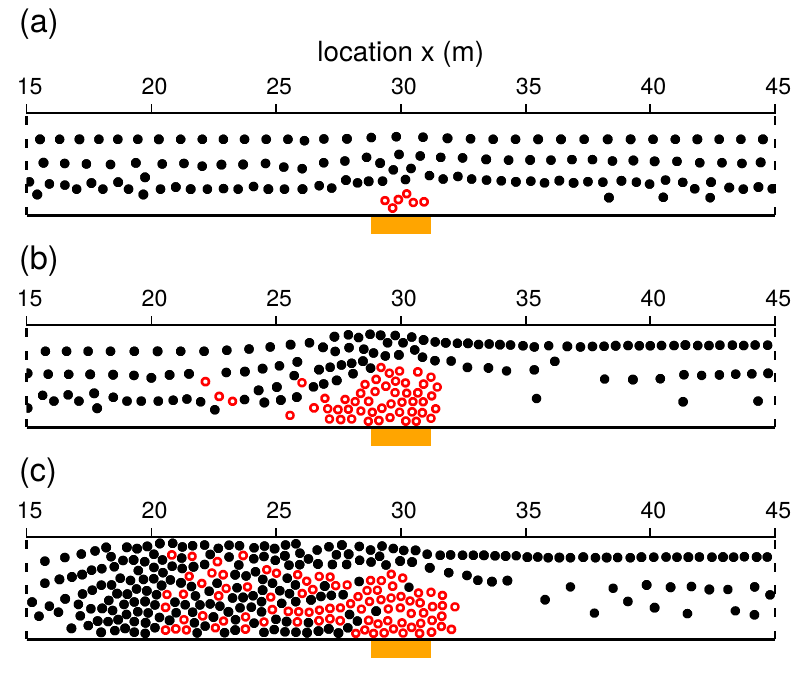}\vspace{-0.3cm}
	\caption{(Color online) Same as Fig.~\ref{fig:snapshot_bi}, but for unidirectional flow. Passersby are walking from the left to the right. We can observe (a) free flow phase in case of $Q = 5$~P/s and $s = 0.5$, which is similar to the case of bidirectional flow, (b) localized jam phase in case of $Q = 5$~P/s and $s = 1$, in which passersby walk slow near the attraction before they pass the area, and (c) extended jam phase in case of $Q = 5$~P/s and $s = 1.8$, in which the passersby queue is growing towards the left boundary. Sometimes the extended jam phase ends up in freezing phenomena if $Q$ and $s$ are very large.}
	\label{fig:snapshot_uni} 
\end{figure}

In unidirectional flow, we can also define the free flow phase if $Q$ and $s$ are small [see Fig.~\ref{fig:snapshot_uni}(a)]. The localized jam phase appears in the vicinity of the attraction for medium and high $Q$ with the intermediate range of $s$, as can be seen from Fig.~\ref{fig:snapshot_uni}(b). Passersby walk slow near the attraction because of reduced walking area, and then they recover their speed after walking away from the attraction. One can observe that pedestrians walking away from the attraction tend to form lanes. This is possible because the standard deviation of speed among the walking away pedestrians is not significant after the pedestrians recover their speed. According to the study of Moussa\"{i}d~\textit{et al.}~\cite{Moussaid_PLOS2012}, the formation of pedestrian lanes is stable when pedestrians are walking at nearly the same speed. Once the walking away pedestrians form lanes, the lanes are not likely to collapse. An extended jam phase can be observed when both $Q$ and $s$ are large, in which the pedestrian queue is growing towards the left boundary and then the queue is persisting for a long period of time [see Fig.~\ref{fig:snapshot_uni}(c)]. In the extended jam phase, the attendee cluster does not maintain its semi-circular shape any more in that passersby seize up the attracted pedestrians. Meanwhile, pedestrians in the queue still can slowly walk towards the right side of the corridor as they initially intended. When $Q$ and $s$ are very large, the extended jam phase can end up in freezing phenomena with a certain probability, indicating that passersby cannot proceed beyond the attraction due to the clogging effect. Some passersby are pushed out towards the attraction by the attracted pedestrians and inevitably they prevent attracted pedestrians from joining the attraction. Consequently, the attracted pedestrians cannot approach the boundary of the attendee cluster although they keep their walking direction towards the attraction. Simultaneously, the passersby near the attraction attempt to walk away from the attraction, but they cannot because they are blocked by the attracted pedestrians. Eventually, the pedestrian movements near the attraction come to a halt.

In order to reflect speed variation among passersby for a given time $t$ and segment $x$, we introduce the local standard deviation $\sigma(x, t)$, which is given as
\begin{eqnarray} \label{eq:efficiency_sd}
	\sigma(x, t) = \sqrt{\frac{1}{|N(x, t)|} \sum_{i\in N(x, t)}\left[ E_i -E(x,t) \right]^{2}}.
\end{eqnarray}
If $\sigma(x, t)$ is $0$, the speed of passersby is homogeneous in segment $x$ for a given time $t$. On the other hand, large $\sigma(x, t)$ indicates significant speed difference among passersby, and possibly suggests the existence of stop-and-go motions in passerby flow with low local efficiency~\cite{Chraibi_PRE2015, Dietrich_PRE2014, Tordeux_JPhysA2016}.

\begin{figure}[!t]
	\centering
	\includegraphics[width=\columnwidth]{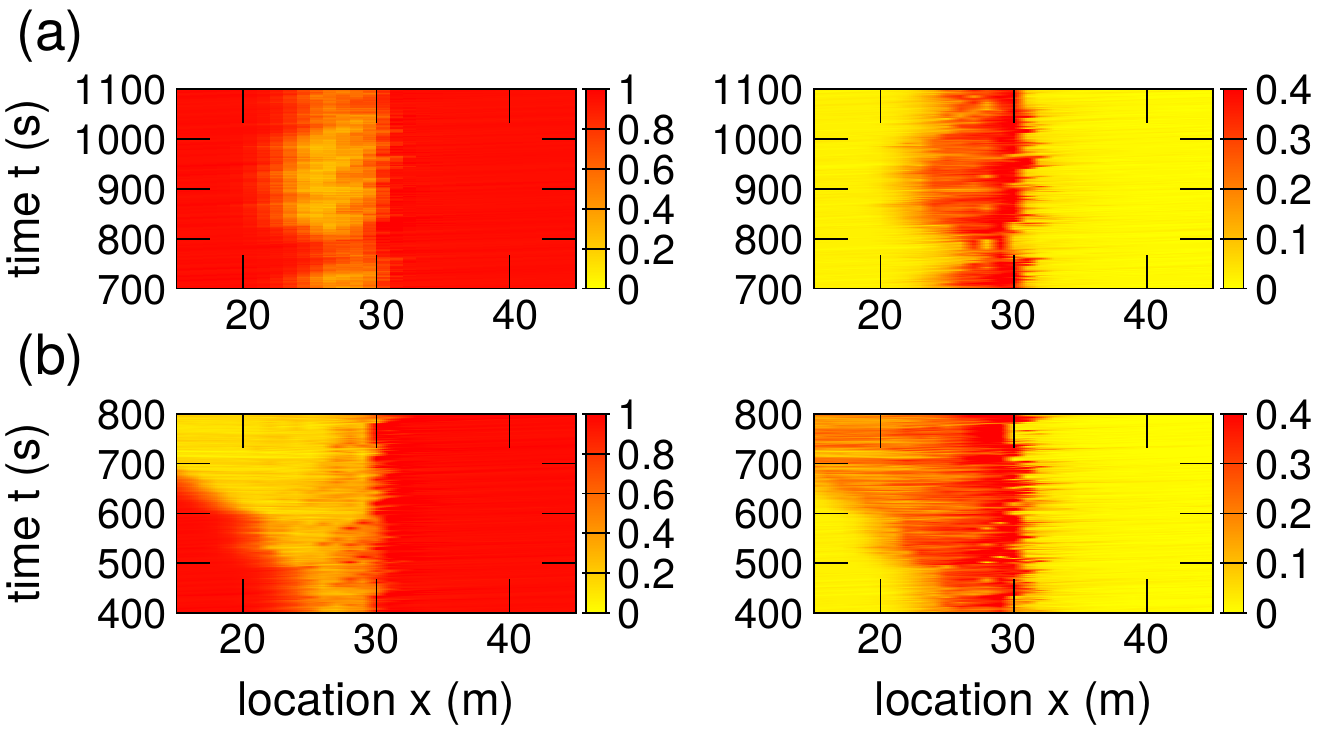}
	\caption{(Color online) Plots of $E(x, t)$ (left column) and $\sigma(x, t)$ (right column) for different pedestrian jam patterns for unidirectional flow. (a) Localized jam phase in unidirectional flow with $Q = 5$~P/s and $s = 1$, in which passersby are likely to reduce their speed near the attraction and then speed up when they walk away from the attraction. The local standard deviation $\sigma(x, t)$ is notable in the low efficiency area, reflecting that some passersby walk with high speed while others walk slowly. (b) Extended jam phase in unidirectional flow with $Q = 5$~P/s and $s = 1.8$. The low efficiency area begins to expand towards the left at near $t = 500$~s, and then the local efficiency remains low for a long period of time over a spatially extended area.}
	\label{fig:spatiotemporal_uni}
\end{figure}

\begin{figure}[!t]
	\centering
	\includegraphics[width=.8\columnwidth]{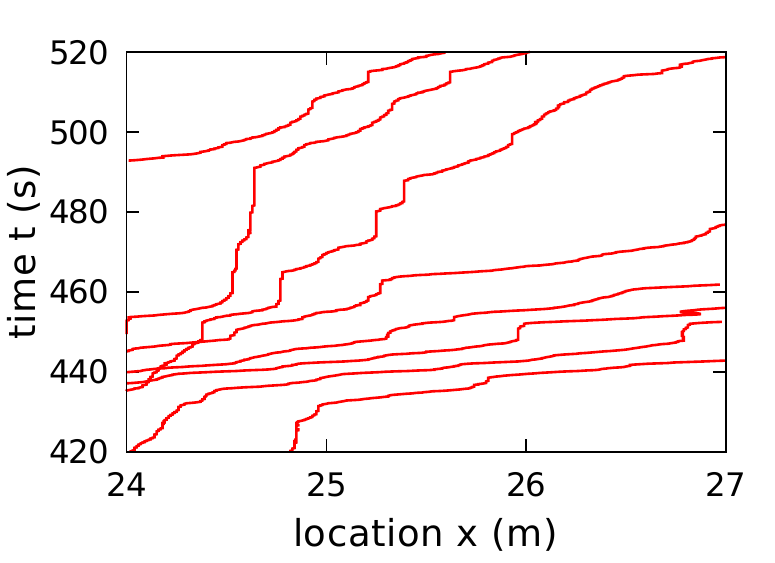}
	\caption{(Color online) Representative trajectories of passersby in the extended jam phase ($Q = 5$~P/s and $s = 1.8$). One can observe that the progress of the passersby is interrupted several times, indicating stop-and-go motions near the attraction.}
	\label{fig:stop-and-go}
\end{figure}

As can be seen from Fig.~\ref{fig:spatiotemporal_uni}(a), in the localized jam phase, one can observe a low-efficiency area in the vicinity of the attraction. In the low-efficiency area, passersby are likely to reduce their speed near the attraction and then speed up when they walk away from the attraction. In the extended jam phase, the low-efficiency area appears near the attraction as in the localized jam [see Fig.~\ref{fig:spatiotemporal_uni}(b)]. In contrast to the case of localized jam, the low efficiency area begins to extend towards the left boundary and then the local efficiency remains low for a long period of time over a spatially extended area. In the low-efficiency area, some passersby move while others are at near standstill, and consequently stop-and-go motions can be observed as reported in previous studies~\cite{Chraibi_PRE2015, Dietrich_PRE2014, Tordeux_JPhysA2016}. Figure~\ref{fig:stop-and-go} shows the presence of stop-and-go motions near the attraction. Therefore, the local standard deviation $\sigma(x, t)$ becomes notable in the low-efficiency area. Due to the passersby flowing out from the low efficiency area, a high local standard deviation $\sigma(x, t)$ is observed near the attraction. Although the average speed of passersby is significantly decreased in the left part of the corridor, the local standard deviation $\sigma(x, t)$ is not zero, indicating that the speed variation among passersby is still observable.

\begin{figure}[!t]
	\centering
	\begin{tabular}{cc}
		\includegraphics[width=.49\columnwidth]{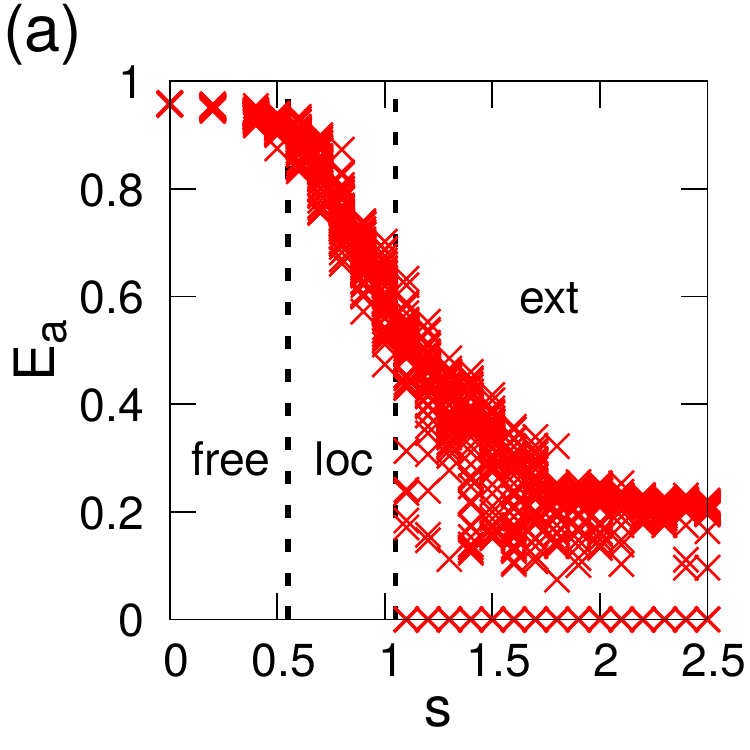}\hspace{-0.3cm}&
		\includegraphics[width=.49\columnwidth]{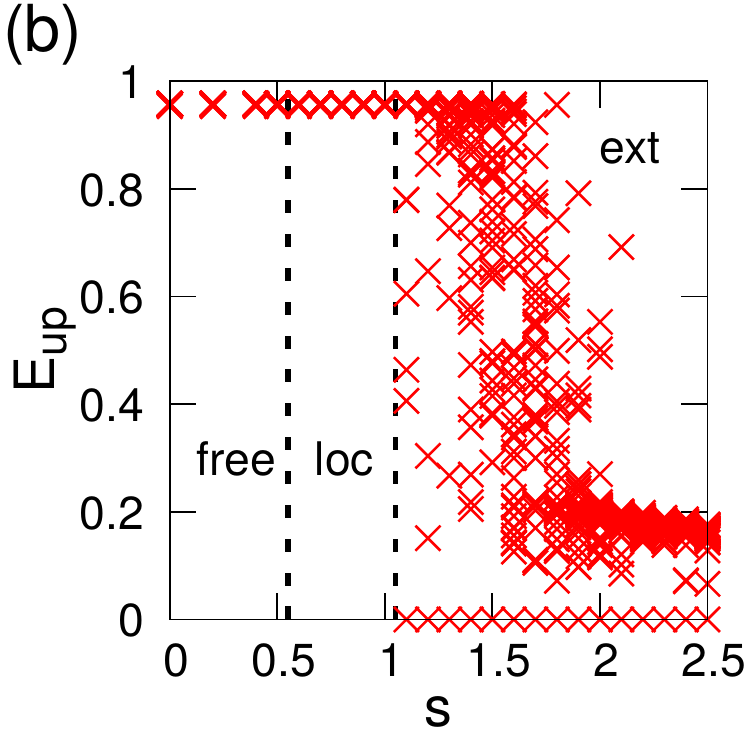}\\
		\includegraphics[width=.49\columnwidth]{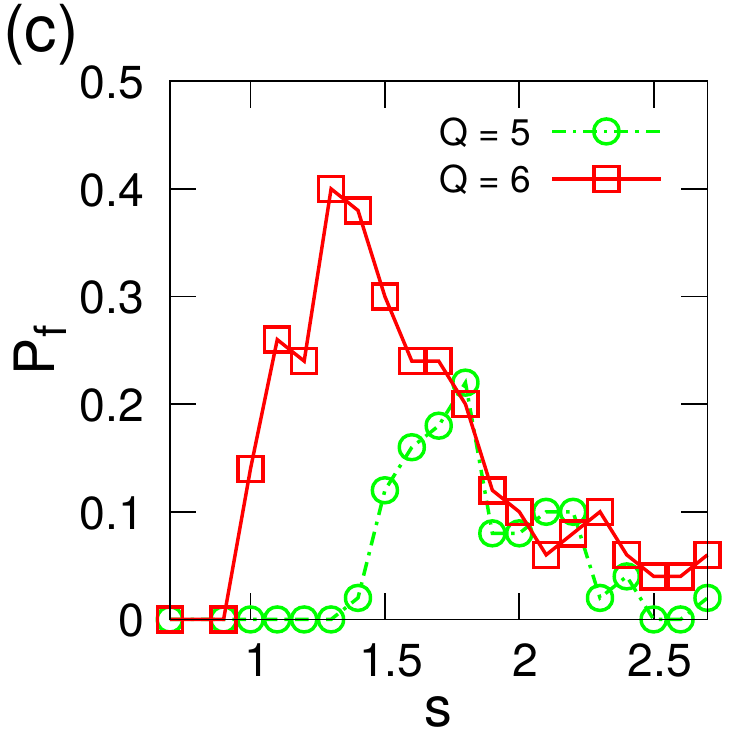}\hspace{-0.3cm}&
		\includegraphics[width=.49\columnwidth]{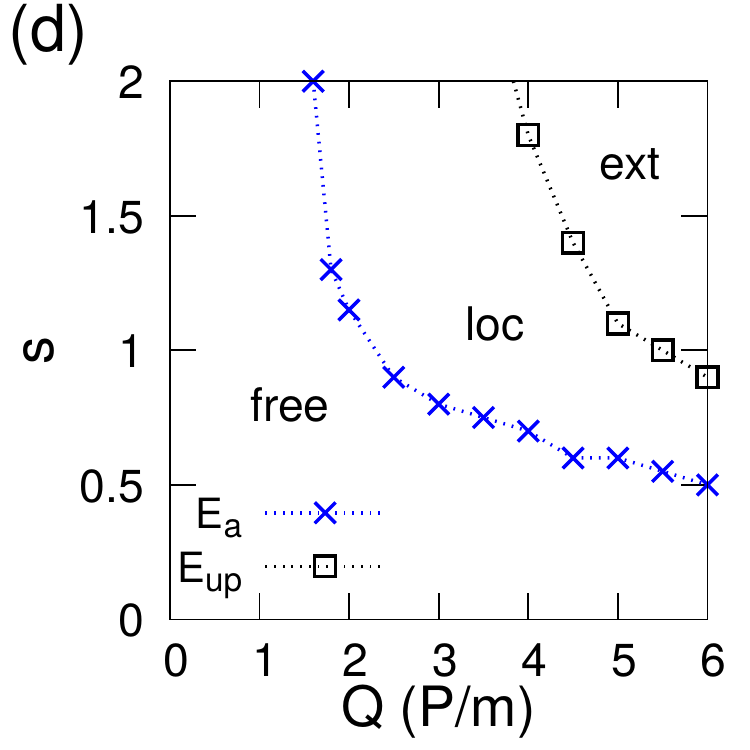}\vspace{-0.3cm}
	\end{tabular}	
	\caption{(Color online) (a) Stationary state average of local efficiency near the attraction $E_{a}$ against $s$ in bidirectional flow with $Q = 5$~P/s. Each point depicts a value obtained from one simulation run. Here, free, loc, and ext indicate the free flow phase, the localized jam phase, and the extended jam phase, respectively. In the extended jam phase, one can observe freezing phenomena with a certain probability $P_f$. (b) Same as (a), but for $E_{up}$, measured upstream of the attraction. (c) Freezing probability $P_f$ as a function of influx $Q$ and social influence parameter $s$ in unidirectional flow. (d) Phase diagram summarizing the numerical results of unidirectional flow. The parameter space of pedestrian influx $Q$ and social influence parameter $s$ is divided into different phases by means of local efficiency measures.}
	\label{fig:PhaseDiagram_uni} 
\end{figure}

Based on observations presented in Figs.~\ref{fig:snapshot_uni} and~\ref{fig:spatiotemporal_uni}, we characterize the localized jam and extended jam phases in terms of $E(x)$ as in Eq.~(\ref{eq:E_x}). Similarly to the bidirectional flow scenario, we select a section of $27\text{~m} \leq x \leq 33\text{~m}$ to calculate $E_{a}$. Likewise, a section of $12\text{~m} \leq x \leq 18\text{~m}$ is selected for upstream of the attraction, and $E_{up}$ denotes the minimum value of $E(x)$ in the section. Figures~\ref{fig:PhaseDiagram_uni}(a) and \ref{fig:PhaseDiagram_uni}(b) provide plots of local efficiency measures in the stationary state, $E_{a}$ and $E_{up}$, produced with $Q = 5$~P/s. For small values of $s$, free flow phase 
can be characterized by 
\begin{eqnarray}\label{eq:uni_free}
	E_{a} \approx 1\ \textrm{and}\ E_{up} \approx 1.
\end{eqnarray}
For $0.6 \leq s \leq 1.05$, data points of $E_{a}$ show a clear decreasing trend against $s$. As can be seen from Fig.~\ref{fig:PhaseDiagram_uni}(b), data points of $E_{up}$ are still near 1 up to $s = 1.05$. Thus, the localized jam phase can be characterized by 
\begin{eqnarray}\label{eq:uni_localized}
	0 < E_{a} < 1\ \textrm{and}\ E_{up} \approx 1.
\end{eqnarray}
When $s$ is larger than 1.05, some data points of $E_{a}$ and $E_{up}$ become zero, indicating that freezing phenomena can be observed. In contrast, positive $E_{a}$ and $E_{up}$ values show a decreasing trend against $s$ for a section of $1.05 \leq s \leq 2$; then they become nearly constant if $s$ is larger than 2. Consequently, the extended jam phase can be characterized by 
\begin{eqnarray}\label{eq:uni_extended}
	0 \leq E_{a} < 1\ \textrm{and}\ 0 \leq E_{up} < 1. 
\end{eqnarray}

In contrast to the bidirectional flow scenario, $P_f$ is always smaller than $1$ in unidirectional flow, indicating that the freezing phase does not exist [see Fig.~\ref{fig:PhaseDiagram_uni}(c)]. However, there exists parameter space producing $P_f>0$, inferring that freezing phenomena can be observed depending on random seeds. Interestingly, in unidirectional flow, $P_f$ is increasing and then decreasing against $s$ for large $Q$. It can be understood that the proportion of passersby decreases considerably as $s$ increases above a certain value, so the attracted pedestrians are less likely blocked by the passersby.

Figure~\ref{fig:PhaseDiagram_uni}(d) summarizes numerical results of phase characterizations. We divide the parameter space of $Q$ and $s$ into different phases by means of local efficiency measures, $E_{a}$ and $E_{up}$. Note that, in the extended jam phase, one can observe freezing phenomena with a certain probability $P_f$.

\subsection{Microscopic understanding of jamming transitions}
\label{subsec:explainJamming}
In previous subsections, we have observed various jam patterns. In bidirectional flow, the free flow phase can turn into a freezing phase if $Q$ and $s$ are large. Jamming transitions in unidirectional flow are different from those of bidirectional flow: from free flow to localized jam, and then to extended jam phases. In addition, it is possible that the extended jam phase ends up in freezing 
phenomena for large $Q$ and $s$. 

While previous sections focused on describing collective patterns of various jam patterns, this section presents the appearance of such different patterns at the individual level in a unified way. This inspired us to take a closer look at the conflicts among pedestrians. Similar to previous studies~\cite{Kirchner_PRE2003, Nowak_PRE2012}, we employ a conflict index to measure the average number of conflicts per passerby. When two pedestrians are in contact and hinder each other, we call this situation a conflict. The number of conflicts $N_{c, i}(t)$ is evaluated by counting the number of pedestrians who hinder the progress of passerby $i$ at time $t$. In our simulations, most conflicts appear near the attraction; therefore, we calculate the conflict index for pedestrians in location $x$ such that $25\text{~m} \leq x \leq 35\text{~m}$. The conflict index is measured as
\begin{eqnarray}\label{eq:ConflictIndex}
	n_c(t) = \frac{1}{|N_p|} \sum_{i\in N_p}{N_{c, i}(t)},
\end{eqnarray}
where $N_p$ is the set of passersby near the attraction. 

\begin{figure}[!t]
	\centering
	\includegraphics[width=\columnwidth]{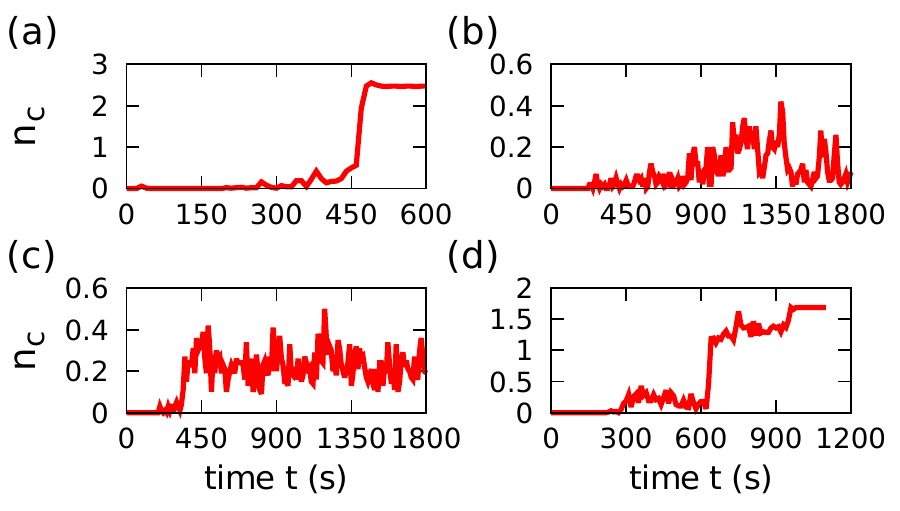}\vspace{-0.3cm}
	\caption{(Color online) Representative time series of conflict index $n_c(t)$. (a) In the beginning, the $n_c(t)$ curve of the freezing phase in bidirectional flow with $Q = 4$~P/s and $s = 1$ is virtually zero, indicating that the pedestrian flow is initially free flow. However, the curve shows a sharp increase at near $t = 460$~s, signaling the onset of freezing phenomenon. (b) The $n_c(t)$ curve of the localized jam phase in unidirectional flow with $Q = 5$~P/s and $s = 1$ exhibits an upward trend until $t = 1350$~s, and then it shows a downward trend leading to zero. (c) For the extended jam phase in unidirectional flow with $Q = 5$~P/s and $s = 1.8$, one can see that the $n_c(t)$ curve sharply increases at near $t = 320$~s, and thereafter it oscillates around $n_c(t) = 0.25$. (d) Same parameter combination $(Q, s)$ as (c), but with a different set of random seeds. The behavior of the $n_c(t)$ curve is similar to that of the extended jam phase until $t = 600$~s. However, the curve abruptly increases at near $t = 600$~s, indicating the appearance of the freezing phenomenon.}
	\label{fig:conflict}
\end{figure}

The representative time series of conflict index $n_c(t)$ are presented in Fig.~\ref{fig:conflict}. As can be seen from Fig.~\ref{fig:conflict}(a), a sharp increase of the conflict index indicates the appearance of the freezing phenomenon, which leads pedestrian flow into the freezing phase. In Fig.~\ref{fig:conflict}(b), the conflict index increases and then decreases in the course of time. We can observe a localized jam phase in which the jam near the attraction does not further grow upstream. Figures~\ref{fig:conflict}(c) and \ref{fig:conflict}(d) are generated with the same parameter combination $(Q, s) = (5, 1.8)$ in unidirectional flow but with different sets of random seeds. As shown in Fig.~\ref{fig:conflict}(c), in the extended jam phase, the conflict index $n_c(t)$ is maintained near a certain level after reaching the stationary state, indicating the persistent jam in the corridor. In Fig.~\ref{fig:conflict}(d), the behavior of the $n_c(t)$ curve is similar to that of the extended jam phase in the beginning, but the curve abruptly increases at near $t = 600$~s. That is, the pedestrian flow eventually ends up in a freezing phenomenon in that conflicting pedestrians fail to coordinate their movements. 

Apart from the conflict index $n_c(t)$, we have also observed that increasing $s$ for a given value of $Q$ likely increases the size of the attendee cluster and consequently reduces the available walking space near the attraction. The narrower walking space tends to yield higher freezing probability in that pedestrians tend to have less space for resolving conflicts among them. Therefore, we suggest that an attendee cluster can trigger jamming transitions not only by reducing the available walking space but also by increasing the number of conflicts among pedestrians near the attraction. See Appendix~\ref{sec:jamming_mechanisms} for further discussion of this issue.

Furthermore, changing other simulation parameters can affect the jamming transitions. For instance, increasing $t_d$ often results in a larger attendee cluster near the attraction, activating jamming transitions for lower values of $s$. Larger corridor width $W$ possibly reduces the freezing probability for a given value of $s$ by providing additional space for resolving pedestrian conflicts. However, increasing $W$ does not effectively reduce the freezing probability when $W$ is large enough in that an attendee cluster can grows further as $W$ grows. In Appendix~\ref{sec:sensitivity}, we show the influence of $t_d$ and $W$ on jamming transitions. 

\subsection{Fundamental diagram}

\begin{figure*}[!t]
	\centering	
	\includegraphics[width=\textwidth]{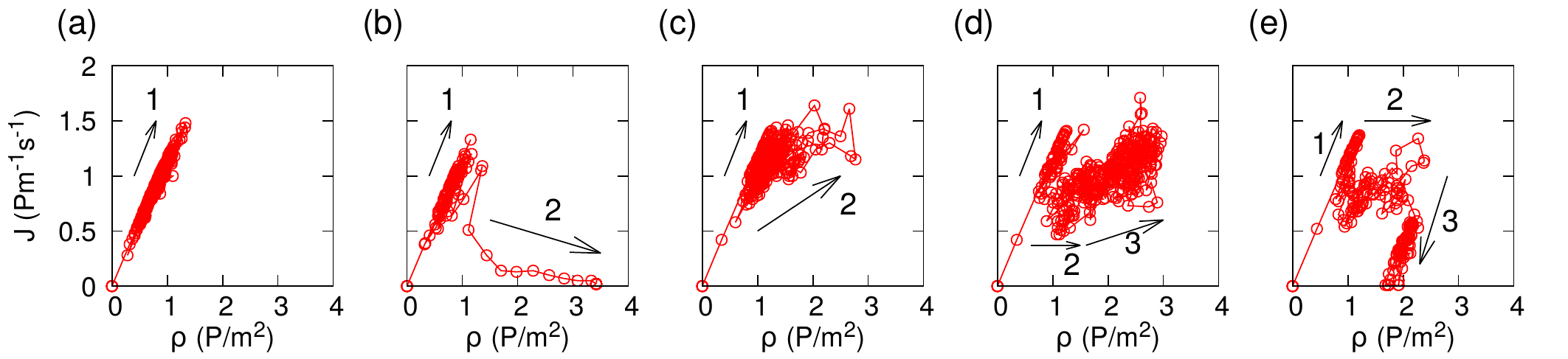}\\
	\includegraphics[width=\textwidth]{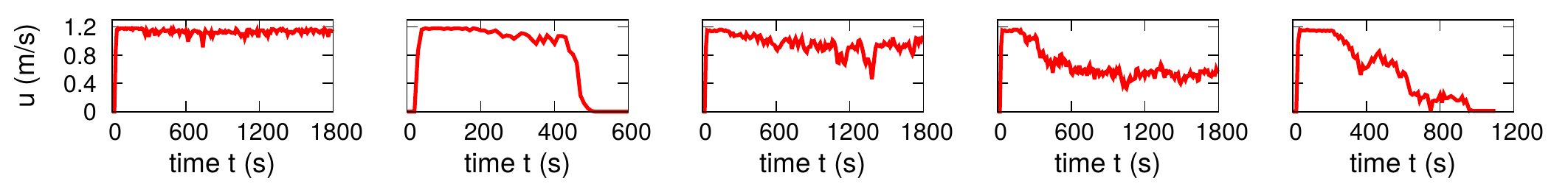}\vspace{-0.2cm}\\
	\caption{(Color online) Fundamental diagram of passerby traffic near the attraction. For each panel, the upper part shows the fundamental diagram and the lower part shows the corresponding local speed $u$. Note that pedestrian flow quantities $J$, $\rho$, and $u$ are calculated for passersby, not including attracted pedestrians. Arrows are guide for the eyes, indicating the evolution of fundamental diagram in the course of time. (a) Free flow phase in bidirectional flow with $Q = 4$~P/s and $s = 0.5$. (b) Freezing phase in bidirectional flow with $Q = 4$~P/s and $s = 1$. (c) Localized jam phase in unidirectional flow with $Q = 5$~P/s and $s = 1$. (d) Extended jam phase in unidirectional flow with $Q = 5$~P/s and $s = 1.8$. (e) Generated from the same parameter combination $(Q, s)$ of as (d), but from a different set of random seeds. The passerby flow turns from an extended jam into the freezing phenomenon.}
	\label{fig:FD} 
\end{figure*}

In addition to the phase diagram presented in Figs.~\ref{fig:PhaseDiagram_bi}(c) and \ref{fig:PhaseDiagram_uni}(d), one can further describe the dynamics of passerby flow by means of a fundamental diagram. The fundamental diagram depicts the relationship between flow and density, which has been widely applied to analyze traffic dynamics to represent various phenomena including hysteresis~\cite{Laval_TrB2011} and capacity drop~\cite{Hall_TRR1991}. We calculate pedestrian flow quantities including density $\rho$, speed $u$, and flow $J$ over two 6-m-long segments in length for every 5 seconds. Note that pedestrian flow quantities $J$, $\rho$, and $u$ are calculated for passersby, not including attracted pedestrians. The details are explained in Appendix~\ref{sec:particle-in-cell}. The first section of $27\text{~m} \leq x \leq 33\text{~m}$ is selected for passerby traffic near the attraction and $12\text{~m} \leq x \leq 18\text{~m}$ for upstream traffic. 

To explore the dynamics of passerby traffic, we plot fundamental diagrams for different phases in bidirectional flow near the attraction. As shown in Fig.~\ref{fig:FD}(a), in the free flow phase, a linear relationship between $\rho$ and $J$ is observed. The corresponding local speed $u$ stays near comfortable walking speed $v_0 = 1.2$~m/s. In Fig.~\ref{fig:FD}(b), an inverse-$\lambda$ shape is observed, reflecting that capacity drop occurs near $\rho = 1.2$~P/$m^{2}$ and then the pedestrian traffic turns into the freezing phase as depicted in Fig.~\ref{fig:spatiotemporal_bi}(b). The corresponding local speed $u$ curve begins to sharply decrease at near $t = 460$~s, relevant to the appearance of the congestion branch. This is similar to the metastable state induced by conflicts among pedestrians~\cite{Ezaki_PRE2012}.

Likewise, we also plot fundamental diagrams and corresponding local speed curves for various jam patterns in unidirectional flow. Figure~\ref{fig:FD}(c) shows that in the localized jam phase, $(\rho, J)$ begins to scatter after the density level reaches around $\rho = 1.2$~P/$m^{2}$. The cluster of scattered data points reflects that speed fluctuation begins to appear, in agreement with Fig.~\ref{fig:spatiotemporal_uni}(b). In contrast, the fundamental diagram upstream of the attraction only shows a linear relationship between $\rho$ and $J$, similar to arrow~1 in Fig.~\ref{fig:FD}(c). That is, upstream traffic is not influenced by the speed reduction near the attraction, thus speed fluctuation is invisible.

In the extended jam phase, a $\mu$ shape is observed in the fundamental diagram as depicted in Fig.~\ref{fig:FD}(d). As indicated by arrow~2, one can see a cluster of $(\rho, J)$ slightly off from the free flow branch, which corresponds to a moderate speed drop from near $t = 320$~s to near $t = 510$~s. Simultaneously, the maximum flow rate $J$ is lower than that in the free flow branch. This can be understood as a transition period in which local speed $u$ is gradually decreasing. After the transition period, one can see data points of $(\rho, J)$ widely spread over in the fundamental diagram while the local speed $u$ slightly oscillates around $u = 0.5$~m/s. 

Figure~\ref{fig:FD}(e) shows fundamental diagrams obtained from the same parameter combination $(Q, s)$ of Fig.~\ref{fig:FD}(d) but from a different set of random seeds. Similar to the case of the extended jam phase, a free flow branch appears, and then one can observe scattered data points of $(\rho, J)$ indicating that the local speed $u$ near the attraction gradually decreases. However, after showing such scattered data points, congestion branches are observed as indicated by the arrow in Fig.~\ref{fig:FD}(e). Interestingly, the behavior of congestion branches observed from Figs.~\ref{fig:FD}(e) and \ref{fig:FD}(b) is different. In Fig.~\ref{fig:FD}(b), the flow $J$ is decreasing as density $\rho$ increases, because there are no outflowing passersby near the attraction while additional pedestrians arrive behind the stopped passersby. On the other hand, the congestion branch in Fig.~\ref{fig:FD}(e) indicate that the flow $J$ is decreasing as density $\rho$ decreases due to passersby flowing out from the attraction. 

\section{Conclusion}
\label{sec:Conclusion}
This study has numerically investigated jamming transitions in pedestrian flow interacting with an attraction. Our simulation model is mainly based on the social force model for pedestrian motions and joining probability reflecting the social influence from other pedestrians. For different values of pedestrian influx $Q$ and the social influence parameter $s$, we characterized various pedestrian flow patterns for bidirectional and unidirectional flows. In the bidirectional flow scenario, we observed a transition from the free flow phase to the freezing phase in which oppositely walking pedestrians reach a complete stop and block each other. However, a different transition behavior appeared in unidirectional flow scenario: from the free flow phase to the localized jam phase, and then to the extended jam phase. One can also see that the extended jam phase end up in freezing phenomena with a certain probability when pedestrian flux is high with strong social influence. It is noted that these results are qualitatively the same for values of simulation time step smaller than 0.05~s, seemingly due to the introduction of an attainable walking speed in Eq.~(\ref{eq:desired_speed}). 

The findings of this study can be interpreted in line with the freezing-by-heating phenomenon observed in particle systems~\cite{Helbing_PRL2000}. Helbing~\textit{et al.}~\cite{Helbing_PRL2000} demonstrated that increasing noise intensity in particle motions leads to the freezing phenomenon, in which particles tend to block each other in a straight corridor. We observed the same phenomenon from pedestrians flow interacting with an attraction. However, it should be noted that Ref.~\cite{Helbing_PRL2000} did not state possible sources of the noise, since that study presented the noise as an abstract concept. Our study suggests that existence of an attraction in pedestrian flow can be a source of such noise. 

Our study highlights that attractive interactions between pedestrians and an attraction can lead to jamming transitions. From the results of numerical simulations, we observed that an attendee cluster can trigger jamming transitions not only by reducing the available walking space but also by increasing the number of conflicts among pedestrians near the attraction. The conflicts arose mainly because attracted pedestrians interfered with passersby who were not interested in the attraction. If the average number of conflicts per passerby is maintained under a certain level, the appearance of freezing phenomena can be prevented. However, when the pedestrian flux is high with strong social influence, the conflicting pedestrians may not be able to have enough time to resolve the conflicts. Therefore, we note that moderating pedestrian flux is important in order to avoid excessive pedestrian jams in pedestrian facilities when the social influence is strong. 

In order to focus on essential features of jamming transitions, this study has considered simple scenarios of pedestrian flow in a straight corridor. Further studies need to be carried out in order to improve the presented models. To mimic pedestrian stopping behavior, pedestrians are represented as non elastic solid disks, indicating that compression among pedestrians is not modeled. The interpersonal friction effect~\cite{Helbing_PRE2007, Yu_PRE2007} needs to be included in the equation of motion for crowd pressure predictions. The joining behavior model in Eq.~(\ref{eq:P_a}) can be further improved and extended by adding additional behavioral features. For instance, explicit representation of group behavior~\cite{Zanlungo_PRE2014} and an interest function~\cite{Kielar_SMPT2016} can be added to the joining behavior model. A natural progression of this work is to analyze the numerical simulation results from the perspective of capacity estimation. Capacity estimation can be performed to calculate the optimal capacity, balancing the mobility needs for passersby and the activity needs for attracted pedestrians. The concept of stochastic capacity~\cite{Minderhoud_TRR1997, Geistefeldt_ISTTT2009, Kerner_PRE2014_vol89} can also be studied as an extension of this study. In this study, for some parameter values, speed breakdown is observed depending on random seeds, inferring that capacity might follow a probability distribution. Future studies can be planned from the perspective of pedestrian flow experiments. Although the joining behavior presented in this study might not be controlled in experimental studies, the experiments can be performed for different levels of pedestrian flux and joining probability. For various experiment configurations, the number of conflicts among pedestrians can be measured and the influence of the conflicts on pedestrian jams can be analyzed. 

\section*{Acknowledgements}
J.K., T.L., and I.K. would like to thank Aalto Energy Efficiency research program (Light Energy-Efficient and Safe Traffic Environments project) for financial support. J.K. is grateful to CSC-IT Center for Science, Finland for providing computational resources. H.-H.J. acknowledges financial support from Basic Science Research Program through the National Research Foundation of Korea grant funded by the Ministry of Education (Grant No. 2015R1D1A1A01058958).

\appendix
\section{Streamline approach for passerby steering behavior}
\label{sec:streamline}
For passerby traffic moving near an attraction, an attendee cluster can act like an obstacle. We assume that passersby set their initial desired walking direction $\vec{e}_{i, 0}$ along the streamlines. As reported in a previous study~\cite{Kwak_PLOS2015}, the shape of an attendee cluster near an attraction can be approximated as a semicircle. By doing so, we can set the streamline function $\psi$ for passerby traffic similar to the case of fluid flow around a circular cylinder in a two dimensional space~\cite{Batchelor_2000, Anderson_2010}:
\begin{eqnarray}\label{eq:streamline}
	\psi = v_0 d_{zA} \sin(\theta_z) \left(1- \frac{r_c}{d_{zA}} \right), %
\end{eqnarray}
where $v_0$ is the comfortable walking speed and $d_{zA}$ is the distance between the center of the semicircle $A$ and location $z = (x, y)$. The angle $\theta_z$ is measured between $y = 0$~m and $\vec{d}_{zA}$. The attendee cluster size at time $t$ is denoted by $r_c = r_c(t)$. To measure $r_c$, we slice the walking area near the attraction into thin layers with the width of a pedestrian size (i.e., $2r_i = 0.4$~m) in the horizontal direction. From the bottom layer to the top one, we count the number of layers consecutively occupied by attendees. The attendee cluster size $r_c$ can then be obtained by multiplying the number of consecutive layers by the layer width $0.4$~m. The initial desired walking direction $\vec{e}_{i, 0}$ can be obtained as 
\begin{eqnarray}\label{eq:e_i0}
	\vec{e}_{i, 0} = \left( \frac{\partial \psi}{\partial y}, -\frac{\partial \psi}{\partial x} \right).
\end{eqnarray}
Note that passersby pursue their initial destination, thus their desired walking direction $\vec{e}_{i}$ is identical to $\vec{e}_{i, 0}$ given in Eq.~(\ref{eq:e_i0}), i.e, $\vec{e}_{i} = \vec{e}_{i, 0}$.

\section{Time-to-collision $T_c$}
\label{sec:T_c}
In line with Refs.~\cite{Moussaid_PNAS2011, Karamouzas_PRL2014}, we assume that pedestrian $i$ predicts time-to-collision $T_c$ with pedestrian $j$ by extending current velocities of pedestrians $i$ and $j$, $v_i$ and $v_j$, from their current positions, $x_i$ and $x_j$:
\begin{eqnarray}\label{eq:T_c}
	T_c = \frac{\beta - \sqrt{\beta^2 - \alpha\gamma}}{\alpha},
\end{eqnarray}
where $\alpha = \left\| \vec{v}_i - \vec{v}_j \right\|^2$, $\beta = (\vec{x}_i-\vec{x}_j)\cdot(\vec{v}_i-\vec{v}_j)$, and $\gamma = \left\| \vec{x}_i-\vec{x}_j \right\|^2 - (r_i + r_j)^2$. Note that $T_c$ is valid for $T_c > 0$, meaning that pedestrians $i$ and $j$ are in a course of collision, whereas $T_c < 0$ implies the opposite case. If $T_c = 0$, the disks of pedestrians $i$ and $j$ are in contact.

\section{Numerical integration of Eq.~(\ref{eq:EoM})}
\label{sec:numerical_integration}
Based on the first-order Euler method, the numerical integration of Eq.~(\ref{eq:EoM}) is discretized as  
\begin{equation}\label{eq:Euler_method}
	\begin{split}
		\vec{v}_i(t + \Delta t) &= \vec{v}_i(t) + \vec{a}_i(t)\Delta t, \\
		\vec{x}_i(t + \Delta t) &= \vec{x}_i(t) + \vec{v}_i(t + \Delta t)\Delta t.
	\end{split}	
\end{equation}
Here, $\vec{a}_i(t)$ is the acceleration of pedestrian $i$ at time $t$ and the velocity of pedestrian $i$ at time $t$ is given as $\vec{v}_i(t)$. The position of pedestrian $i$ at time $t$ is denoted by $\vec{x}_i(t)$. 

\section{Discussion of jamming mechanisms}
\label{sec:jamming_mechanisms}
In Sec.~\ref{subsec:explainJamming}, we discussed the dynamics of jamming transitions mainly based on the conflict index [see Eq.~(\ref{eq:ConflictIndex})]. This appendix provides further details of jamming mechanisms. 

In both bidirectional and unidirectional flows, attracted pedestrians often trigger conflicts among pedestrians. When the attracted pedestrians are walking towards the attraction, sometimes they cross the paths of passersby and hinder their walking. Furthermore, such crossing behavior of attracted pedestrians makes others change their walking directions due to the interpersonal repulsion, possibly giving rise to conflicts among the others. Once a couple of pedestrians hinder each other, they need some time and space to resolve the conflict by adjusting their walking directions. If there is not enough space for the movement, the conflict situation cannot be resolved and turns into a blockage in pedestrian flow. Higher pedestrian flux $Q$ can be interpreted as the conflicting pedestrians likely have less time for resolving the conflict, in that additional pedestrians arrive behind the blockage. Once the arriving pedestrians stand behind the blockage, the number of conflicts among pedestrians is rapidly increasing as indicated in Figs.~\ref{fig:conflict}(a) and \ref{fig:conflict}(d). In the case of bidirectional flow, this freezing phenomenon is similar to the freezing-by-heating phenomenon~\cite{Helbing_PRL2000}. We note, however, that the freezing phenomenon in our simulations is caused by attracted pedestrians without noise terms in the equation of motion. 

\begin{figure}[!t]
	\centering
	\includegraphics[width=0.75\columnwidth]{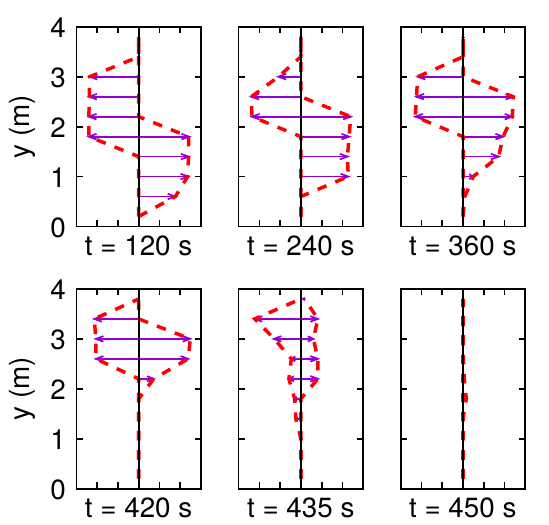}\vspace{-0.3cm}
	\caption{(Color online) Streamwise velocity profiles of passerby traffic in the corridor at $x = 30$~m with $Q = 4$~P/s and $s = 1$ in bidirectional flow. Based on Eq.~(\ref{eq:LocalSpeed}), the velocity vectors are presented for eastbound and westbound passerby traffic in the right and left parts of each panel, respectively. In the course time, the velocity profile curves are gradually shifting upward while their magnitude decreases.}
	\label{fig:VxDistProfile_bi} 
\end{figure}

The onset of freezing phenomena in bidirectional flow can be further understood by looking at streamwise velocity profiles. Figure~\ref{fig:VxDistProfile_bi} visualizes streamwise velocity profiles of passerby traffic in the corridor at $x = 30$~m where the attraction is placed. In the beginning (i.e., $t = 120$~s), one can observe that the velocity vectors of two passerby streams show distinct spatial separation, indicating that passerby lane formation appears to be well maintained. We notice that the velocity magnitude near the bottom of the corridor is virtually zero when the attendee cluster becomes active. In the course of time, the velocity distribution curves shrink while the curves are gradually shifting upward. This implies that two passerby streams moving in opposite directions confront each other, leading to a freezing phenomenon.

\begin{figure}[!t]
	\centering
	\begin{tabular}{cc}
		\includegraphics[width=.49\columnwidth]{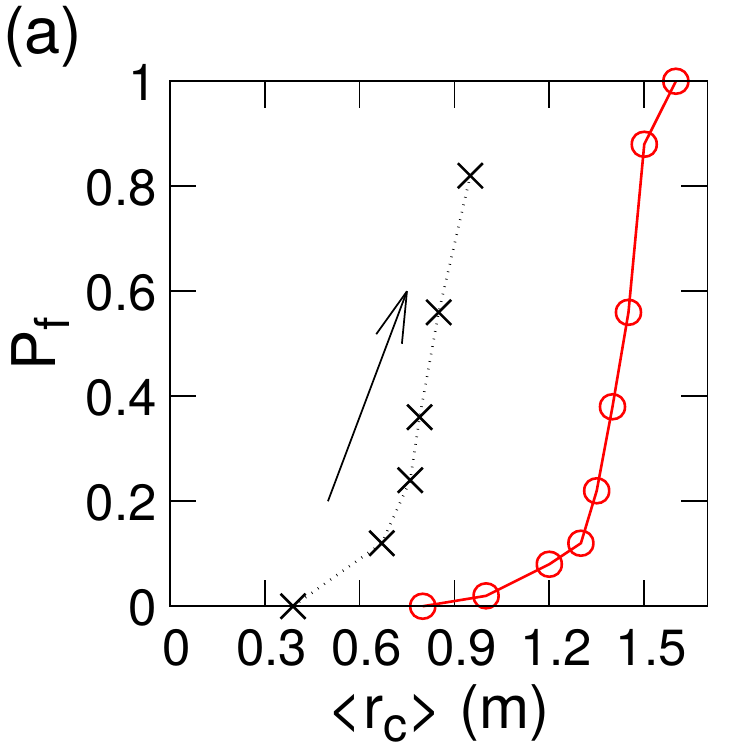}&
		\includegraphics[width=.49\columnwidth]{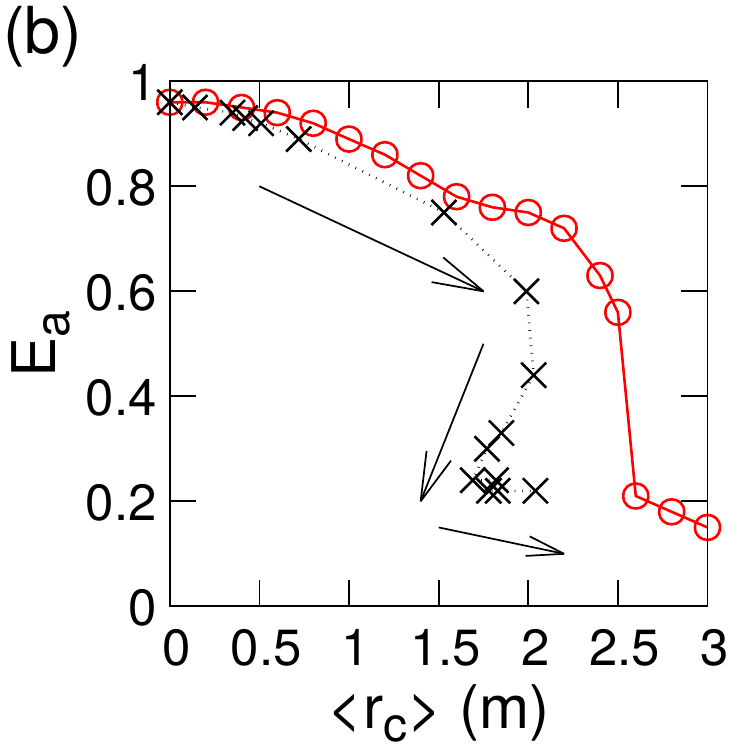}\\
		\includegraphics[width=.49\columnwidth]{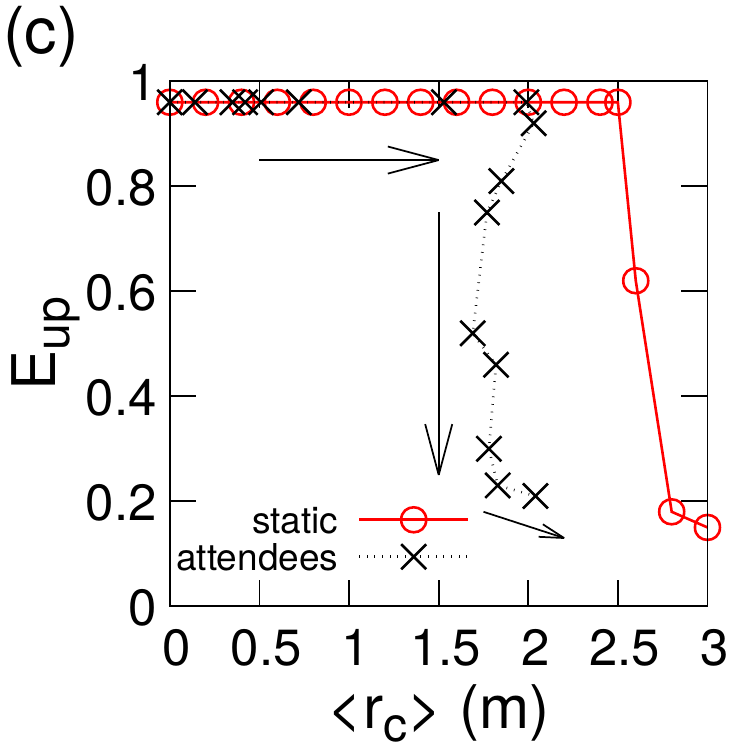}\vspace{-0.2cm}
	\end{tabular}
	\caption{(Color online) Dependence of various measures on the stationary state average of cluster size $\langle r_c \rangle$ for pedestrian influx $Q = 5$~P/s. The results of the static bottleneck and attendee cluster are denoted by $\circ$ and $\times$, respectively. Arrows indicate the direction of increasing $s$, for the results of the attendee cluster. (a) Freezing probability $P_f$ for bidirectional flow indicating the appearance of the freezing phase (b) Local efficiency near the attraction $E_{a}$ for unidirectional flow reflecting the onset of localized jam phase, and (c) local efficiency upstream $E_{up}$ for unidirectional flow, which is relevant to the extended jam phase.}
	\label{fig:compare_static-attendees} 
\end{figure}

Furthermore, the attendee cluster also contributes to jamming transitions by reducing the available space in the corridor. To understand the influence of attendee cluster size on jamming transitions, we perform numerical simulations with a static bottleneck instead of an attraction, in which all the pedestrians are passersby. In doing so, we can exclude the interactions among passersby and attendees, thereby focusing on the influence of reduced available space. For the comparison, we use a stationary state average of cluster size $\langle r_c \rangle$ because the attendee cluster size changes in the course of time but the size of the static bottleneck is constant. In the case of a static bottleneck, we also use the notation of $\langle r_c \rangle$ for convenience. A semicircle with radius $\langle r_c \rangle$ is placed at the center of the lower corridor boundary, acting as a static bottleneck. By changing $\langle r_c \rangle$, we observe the behavior of various measures including $P_f$, $E_{a}$, and $E_{up}$ [see Fig.~\ref{fig:compare_static-attendees}]. 

It is obvious that larger $\langle r_c \rangle$ leads to higher freezing probability $P_f$ for bidirectional flow, as shown in Fig.~\ref{fig:compare_static-attendees}(a). However, $P_f$ of the attendee cluster case is higher than that of a static bottleneck for a given value of $\langle r_c \rangle$. While the $E_{a}$ and $E_{up}$ curves obtained from the static bottleneck case show a clear dependence on $\langle r_c \rangle$, those from attendee cluster do not show clear tendency when $\langle r_c \rangle > 1.5$~m [see Figs.~\ref{fig:compare_static-attendees}(b) and \ref{fig:compare_static-attendees}(c)]. Although increasing $\langle r_c \rangle$ evidently leads to a localized jam transition, it can be suggested that conflicts among pedestrians play an important role in jamming transitions if $\langle r_c \rangle$ is large enough. 

\section{Sensitivity analysis}
\label{sec:sensitivity}
\begin{figure}[!t]
	\centering	
	\begin{tabular}{cc}
		\includegraphics[width=.49\columnwidth]{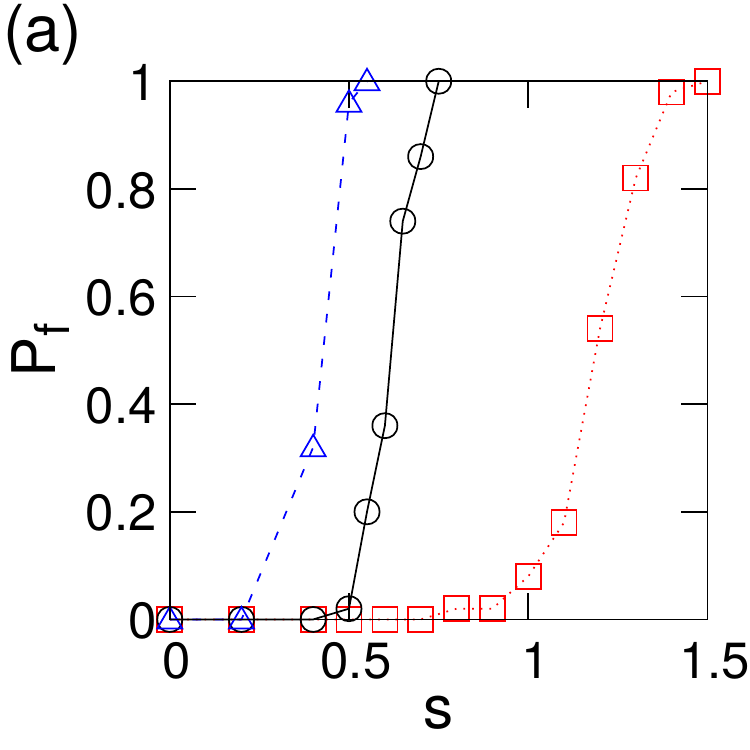}&
		\includegraphics[width=.49\columnwidth]{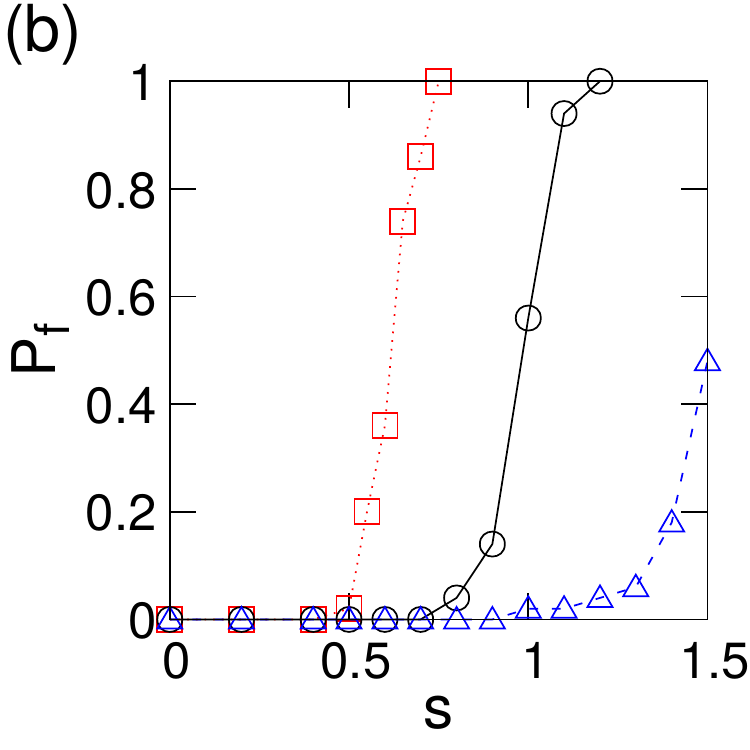}\\
		\includegraphics[width=.49\columnwidth]{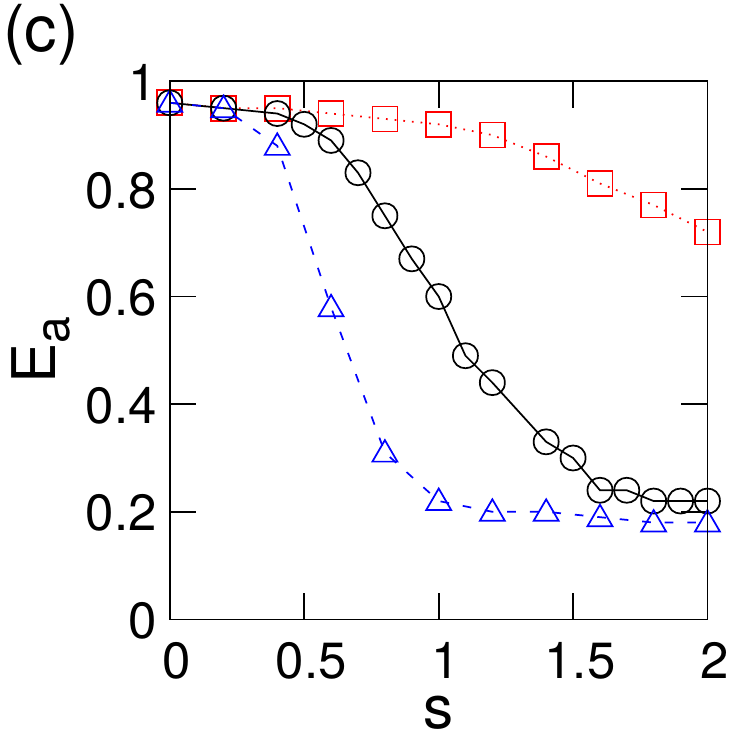}&
		\includegraphics[width=.49\columnwidth]{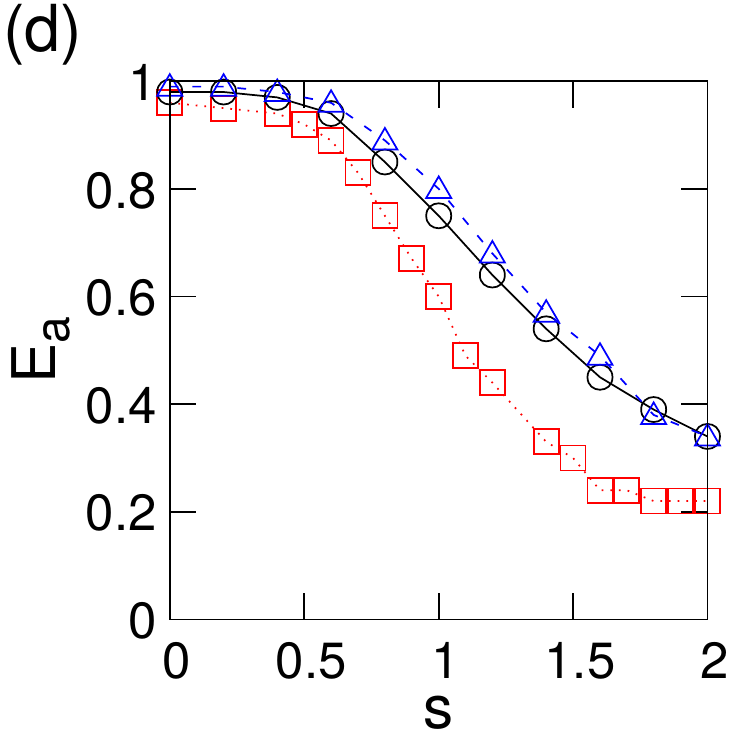}\\
		\includegraphics[width=.49\columnwidth]{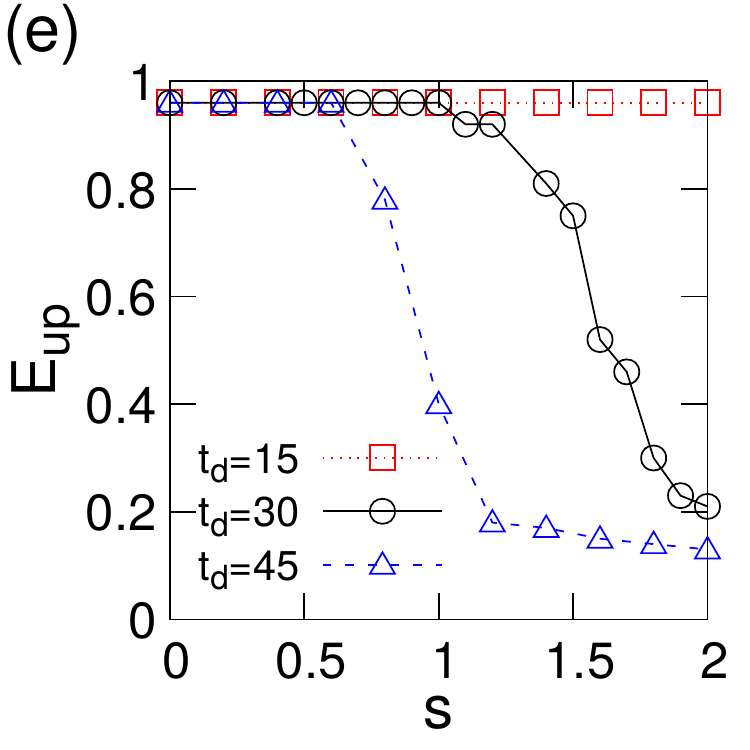}&
		\includegraphics[width=.49\columnwidth]{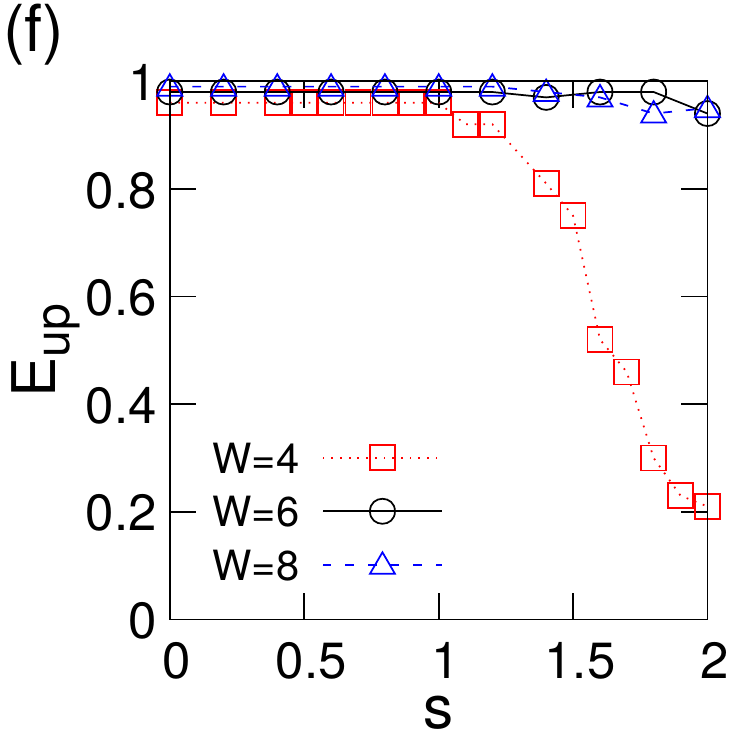}\vspace{-0.3cm}\\
	\end{tabular}
	\caption{(Color online) Dependence of various measures on social influence parameter $s$ for $t_d$ (left column) and $W$ (right column) with pedestrian influx $Q = 5$~P/s. From top to bottom, (a) and (b) Freezing probability $P_f$ for bidirectional flow, (c) and (d) local efficiency near the attraction $E_{a}$ for unidirectional flow, and (e) and (f) local efficiency upstream $E_{up}$ for unidirectional flow. Note that $t_d = 30$~s and $W = 4$~m are the simulation parameter values that we are mainly using in this study.}
	\label{fig:sensitivity} 
\end{figure}

Since the presented results are sensitive to the numerical simulation setup, so we briefly discuss the influence of different simulation parameters especially for the average length of stay $t_d$ and the corridor width $W$. As can be seen from Figs.~\ref{fig:sensitivity}(a), \ref{fig:sensitivity}(c), and \ref{fig:sensitivity}(e) the freezing probably reaches $P_f = 1$ quicker, and local efficiency measures $E_{a}$ and $E_{up}$ tend to decrease faster as $t_d$ grows. One can infer that larger $t_d$ likely leads to having more attendees near attractions, resulting in higher freezing probably and smaller local efficiency measures. Therefore, it can be suggested that increasing $t_d$ activates jamming transitions for lower values of $s$.

In Figs.~\ref{fig:sensitivity}(b), \ref{fig:sensitivity}(d), and \ref{fig:sensitivity}(f), increasing $W$ apparently changes the behavior of $P_f$ in bidirectional flow in that conflicting pedestrians seem to have enough space for resolving the conflicts. Although increasing the corridor width from $W = 4$~m to $W = 6$~m produces notable differences in the $E_{a}$ and $E_{up}$ curves, increasing $W$ further does not seem to yield any significant changes. It is reasonable to suppose that the impact of increasing $W$ becomes less notable for large $W$, in that increased $W$ allows conflicting pedestrians to have more space for resolving the conflicts but also the attendee cluster to grow larger.

\section{Pedestrian flow quantities}
\label{sec:particle-in-cell}
\begin{figure}[!t]
	\centering
	\includegraphics[width=5cm]{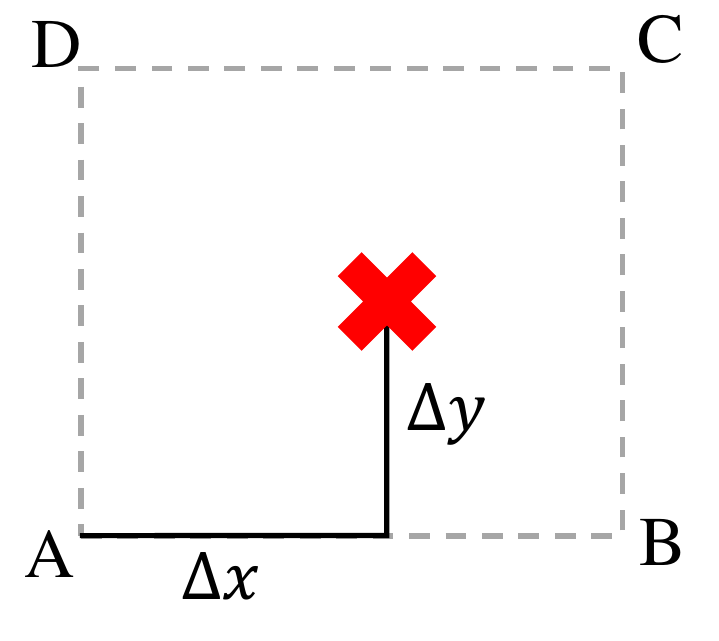} 
	\caption{(Color online) Schematic representation of the grid structure. Points A, B, C, and D indicate the neighboring grid points of pedestrian $i$. The location of pedestrian $i$ is denoted by $\times$. The relative coordinates of pedestrian $i$ with respect to point A in the horizontal and vertical directions are indicated by $\Delta x$ and $\Delta y$, respectively.}
	\label{fig:gridpoints} 
\end{figure}

We evaluated pedestrian flow quantities such as local density, local speed, and local flow. Following the idea of the particle-in-cell method~\cite{Harlow_1962, Dawson_RMP1983, Treuille_ACM2006, Narin_ACM2009}, we convert the discrete number of pedestrians into continuous density field values by using a bilinear weight function $w_{iz}$ for each neighboring grid point $z \in \left\{ A, B, C, D \right\}$, 
\begin{equation}\label{eq:ConversionWeight}
	\begin{split}
		w_{iA} &= (l-\Delta x)(l-\Delta y)/{l^2}, \\
		w_{iB} &= \Delta x (l-\Delta y)/{l^2}, \\
		w_{iC} &= \Delta x \Delta y/{l^2}, \\
		w_{iD} &= (l-\Delta x) \Delta y/{l^2},
	\end{split}	
\end{equation}
where $\Delta x$ and $\Delta y$ indicate the relative coordinates from the left bottom cell center $A$ to the location of pedestrian $i$ (see Fig.~\ref{fig:gridpoints}). Although the use of a Gaussian function is a well-established approach in quantifying the local flow characteristics~\cite{Kwak_PRE2013, Helbing_PRE2007, Moussaid_PNAS2011}, it tends to overestimate the local quantities. This is because it takes into account distant pedestrians. Notice that $\sum {w_{iz}} = 1$ indicating that the weight function $w_{iz}$ reflects the density contribution from pedestrian $i$. We choose grid spacing $l = 2r_i$ on the order of pedestrian size. With the weight function $w_{iz}$, the local density $\rho(\vec{z}, t)$ is defined as 
\begin{eqnarray}\label{eq:LocalDensity}
	\rho(\vec{z}, t) = \sum_{i}^{ } \frac{w_{iz}}{l^2}.	
\end{eqnarray}
Likewise, the local speed $u(\vec{z}, t)$ is given as 
\begin{eqnarray}\label{eq:LocalSpeed}
	u(\vec{z}, t) = \frac{\sum_{i}^{ }{\|\vec{v}_{i}\| w_{iz} }}{\sum_{i}^{ }{w_{iz}}}.
\end{eqnarray}
We can calculate the local pedestrian flow $J(\vec{z}, t)$ as a product of local density and local speed, 
\begin{eqnarray}\label{eq:LocalFlow}
	J(\vec{z}, t) = \rho(\vec{z}, t) u(\vec{z}, t).
\end{eqnarray}

\clearpage
\newpage

\end{document}